\documentstyle[12pt]{article}

\newcommand{\sect}[1]{\setcounter{equation}{0}\section{#1}}

\newcommand{\app}[1]{\setcounter{section}{0}
\setcounter{equation}{0} \renewcommand{\thesection}{\Alph{section}}
\section{#1}}
\newcommand{\eq}{\begin{equation}}
\newcommand{\eqa}{\begin{eqnarray}}
\newcommand{\en}{\end{equation}}
\newcommand{\ena}{\end{eqnarray}}
\newcommand{\enn}{\nonumber \end{equation}}

\def\sk{\vskip .4cm}
\def\noi{\noindent}
\def\om{\omega}
\def\al{\alpha}
\def\be{\beta}
\def\ga{\gamma}
\def\Ga{\Gamma}
\def\del{\delta}

\def\rinv{{1\over {r-r^{-1}}}}

\def\Cb{\bar{C}}

\def\fb{\bar{f}}
\def\gb{\bar{g}}

\def\rhop{{\rho}^{\prime}}

\def\thetap{{\theta}^{\prime}}

\def\onehalf{{1 \over 2}}

\def\epsi{\varepsilon}
\def\we{\wedge}

\def\de{\delta}

\def\part{\partial}

\def\R#1#2{ R^{#1}_{~~~#2} }
\def\Rp#1#2{ (R^+)^{#1}_{~~~#2} }

\def\Rm#1#2{ (R^-)^{#1}_{~~~#2} }
\def\Rinv#1#2{ (R^{-1})^{#1}_{~~~#2} }
\def\Rsecondinv#1#2{ (R^{\sim 1})^{#1}_{~~~#2} }
\def\Rinvsecondinv#1#2{ ((R^{-1})^{\sim 1})^{#1}_{~~~#2} }

\def\Rpm#1#2{(R^{\pm})^{#1}_{~~~#2} }

\def\Rh{{\hat R}}

\def\Rhat#1#2{ \Rh^{#1}_{~~~#2} }
\def\Rbar#1#2{ {\bar R}^{#1}_{~~~#2} }
\def\L#1#2{ \La^{#1}_{~~~#2} }
\def\Linv#1#2{ \La^{-1~#1}_{~~~~~#2} }

\def\Z#1#2{ Z^{#1}_{~~~#2} }

\def\La{\Lambda}

\def\ff#1#2#3{f_{#1~~~#3}^{~#2}}
\def\MM#1#2#3{M^{#1~~~#3}_{~#2}}
\def\cchi#1#2{\chi^{#1}_{~#2}}
\def\ome#1#2{\om_{#1}^{~#2}}
\def\RRhat#1#2#3#4#5#6#7#8{\La^{~#2~#4}_{#1~#3}|^{#5~#7}_{~#6~#8}}
\def\RRhatinv#1#2#3#4#5#6#7#8{(\La^{-1})^
{~#2~#4}_{#1~#3}|^{#5~#7}_{~#6~#8}}
\def\LL#1#2#3#4#5#6#7#8{\La^{~#2~#4}_{#1~#3}|^{#5~#7}_{~#6~#8}}

\def\Cb{{\bf C}}
\def\CC#1#2#3#4#5#6{\Cb_{~#2~#4}^{#1~#3}|_{#5}^{~#6}}
\def\cc#1#2#3#4#5#6{C_{~#2~#4}^{#1~#3}|_{#5}^{~#6}}

\def\C#1#2{ {\bf C}_{#1}^{~~~#2} }
\def\c#1#2{ C_{#1}^{~~~#2} }

\def\Dmat#1#2{D^{#1}_{~#2}}
\def\Dmatinv#1#2{(D^{-1})^{#1}_{~#2}}
\def\DR{\Delta_R}
\def\DL{\Delta_L}
\def\f#1#2{ f^{#1}_{~~#2} }

\def\T#1#2{ T^{#1}_{~~#2} }
\def\Ti#1#2{ (T^{-1})^{#1}_{~~#2} }

\def\M#1#2{ M_{#1}^{~#2} }
\def\qm{q^{-1}}
\def\rminus{r^{-1}}

\def\D{\Delta}
\def\DN{\Delta_{N+1}}
\def\kN{\kappa_{N+1}}
\def\eN{\epsi_{N+1}}

\def\Dp{\Delta^{\prime}}

\def\ep{\epsi^{\prime}}
\def\kp{\kappa^{\prime}}
\def\kpm{\kappa^{\prime -1}}
\def\kpsq{\kappa^{\prime 2}}
\def\km{\kappa^{-1}}

\def\rone{r \rightarrow 1}
\def\qrone{q,r \rightarrow 1}

\def\GLqrN{GL_{q,r}(N)}
\def\IGLqrN{IGL_{q,r}(N)}
\def\IGLqrtwo{IGL_{q,r}(2)}
\def\GLqrNo{GL_{q,r}(N+1)}

\def\UglqrN{U(gl_{q,r}(N))}

\def\Lpm#1#2{L^{\pm #1}_{~~~#2}}
\def\Lmp#1#2{L^{\mp#1}_{~~~#2}}
\def\LLpm{L^{\pm}}
\def\LLmp{L^{\mp}}
\def\LLp{L^{+}}
\def\LLm{L^{-}}
\def\Lp#1#2{L^{+ #1}_{~~~#2}}
\def\Lm#1#2{L^{- #1}_{~~~#2}}

\def\ronelim{\stackrel{r \rightarrow 1}{\longrightarrow}}
\def\qronelim{\stackrel{q=r \rightarrow 1}{\longrightarrow}}

\def\detq{{\det}}
\def\detqr{{\det}}
\def\detqrm{{\det} }
\def\detqrTAB{{\det} \T{A}{B}}
\def\detqrTab{{\det} \T{a}{b}}
\def\P{P}
\def\Qt{Q}
\def\chit{{\partial}}

\def\pp#1#2{\Pi_{#1}^{(#2)}}

\begin{document}

\begin{titlepage}
\rightline{DFTT-9/94}
\rightline{July 1994}
\vskip 2em
\begin{center}{\bf  INHOMOGENEOUS QUANTUM GROUPS $IGL_{q,r}(N)$ :}
{\bf UNIVERSAL ENVELOPING ALGEBRA AND DIFFERENTIAL CALCULUS}
\\[3em]
Paolo Aschieri \\[1em]
{\sl Scuola Normale Superiore\\
Piazza dei Cavalieri 7,  56100 Pisa \\and\\Istituto Nazionale
di Fisica Nucleare,\\ Sezione di Pisa, Italy}\\[3em]
Leonardo Castellani\\[1em]
{\sl Istituto Nazionale di
Fisica Nucleare\\
and\\Dipartimento di Fisica Teorica\\
Via P. Giuria 1, 10125 Torino, Italy.}  \\[3em]
\end{center}

\begin{abstract}
A  review of the multiparametric
linear quantum group $\GLqrN$, its real forms, its
dual algebra $\UglqrN$ and its
bicovariant differential calculus is given in the
first part of the paper.

We then construct the (multiparametric) linear
inhomogeneous quantum group $\IGLqrN$ as a projection from
 $\GLqrNo$, or equivalently, as a quotient of
$\GLqrNo$ with respect to a suitable
Hopf algebra ideal.

A bicovariant
differential calculus
on $IGL_{q,r}(N)$ is explicitly obtained as a
projection from the one on $\GLqrNo$.

Our procedure unifies in a single structure the quantum
plane coordinates and the $q$-group matrix elements $\T{a}{b}$,
and allows to deduce without effort the differential
 calculus on the
$q$-plane $\IGLqrN / \GLqrN$.

 The general theory is illustrated on the example
of $IGL_{q,r}(2)$.

\end{abstract}

\vskip .5cm

\noi DFTT-9/94

\noi  July 1994~~~~~~~~~~~~~~~~~~~~~~~~~~~~~~~~~~~~~~~
{}~~~~~~~~~~~~~~~~~~~~~~~~~~~~~~~~~hep-th/9408031
\vskip .2cm
\noi \hrule
\vskip.2cm
\hbox{\vbox{\hbox{{\small{\it email addresses:}}}\hbox{}}
 \vbox{\hbox{{\small decnet=31890::aschieri,
31890::castellani}}
\hbox{{\small internet = aschieri@sns.sns.it,
castellani@to.infn.it
}}}}

\end{titlepage}
\newpage
\setcounter{page}{1}

\sect{Introduction}

Quantum deformations of inhomogeneous Lie groups
have been studied recently \cite{inhom}. An $R$-matrix
approach has been
independently proposed for $IGL_q(N)$
 in ref.s \cite{Rinhom1} and \cite{Rinhom2} ; in
\cite{Rinhom1} the corresponding
universal enveloping algebra is discussed, while in
\cite{Rinhom2}  a
 bicovariant differential calculus for $IGL_q(N)$
is obtained.
In ref. \cite{Rinhom3} the $IGL_q(N)$ Hopf
 algebra and its bicovariant
differential calculus are obtained via a projection from
$GL_q(N+1)$. The same idea was used
in ref.s \cite{Rinhom4} to
find the bicovariant calculus on the
inhomogeneous $q$-groups
of the $B,C,D$ type, and a $q$-deformation
of the Einstein-Cartan lagrangian of gravity
based on $ISO_q(3,1)$.
\sk
In this paper we construct the {\sl multiparametric}
 $\IGLqrN$ quantum
groups and their bicovariant differential
calculus by using the projective
method of \cite{Rinhom3,Rinhom4}. As we found already in
\cite{Rinhom4}, for inhomogeneous $q$-groups it is
essential to consider the most general (multiparametric) case.
Indeed we find that only in some of these deformations  the
dilatation part can be set to the identity.

All the quantities relevant to their
(bicovariant)  differential calculus are given explicitly:
exterior derivatives,  left-invariant one-forms, Cartan-Maurer
equations, tangent vectors and their $q$-Lie algebra and so on.
The method is illustrated in the case of $\IGLqrtwo$:  the
general formulas are applied and tested on this example
(see the Table).

Our framework allows us to construct the differential
geometry of the (multiparametric) quantum plane in a novel
and easy way.  We obtain a generalization of the
$q$-plane of  ref.  \cite{qplane}; this last is recovered
when setting all parameters equal to a single parameter $q$.
\sk
In Section 2 we recall the basics of  the
 linear quantum groups
 (see ref.s \cite{qgroups1} - \cite{Aschieri1}),
and in Section 3 we discuss their duals
in some detail.  In fact, Sections 2 and 3 are a short review of
 the multiparametric deformations of
$GL_{q,r}(N)$, where $q$ indicates a set of parameters $q_i$,
and of  their universal enveloping algebras.
The usual uniparametric case is recovered for $r=q_i=q$.
For references on multiparametric deformations, see
\cite{Schirrmacher,multiparam}.
\sk
The explicit
construction of the bicovariant differential calculus for
$\GLqrN$, in terms of the dual algebra,  is given in Section 4.
Some new results include an inversion formula for the
left-invariant one-forms in terms of derivatives of
the group elements (the $q$-analogue of $\om=g^{-1}dg$).
For a review of the differential geometry
on quantum groups see for ex. \cite{Aschieri1}.
This subject, initiated in
\cite{Wor}, has been actively developed in recent years:
an incomplete list of references can be found in
\cite{Bernard} - \cite{Schupp}.
\sk
In Section 5 we first present
the quantum group  $IGL_{q,r}(N)$ as a Hopf algebra
with given generators, commutation relations and co-structures.
We then reobtain it as the image of a
projection $P$ from $GL_{q,r}(N+1)$, and show how
the ``mother" Hopf algebra of $GL_{q,r}(N+1)$
determines the Hopf algebra structure on $IGL_{q,r}(N)$.
In the language of Hopf algebra
ideals  $IGL_{q,r}(N)$
is seen as the quotient
of $GL_{q,r}(N+1)$
with respect to a suitable Hopf ideal.

The fundamental representation
of $IGL_{q,r}(N)$ contains the $GL_{q,r}(N)$ elements $\T{a}{b}$
and the ``coordinates" $x^a$, as in the classical case;
in addition, there is also an element $u$ playing
the role of a dilatation.
By fixing some of the parameters $q$, we find that
this element $u$ can be made central, and hence consistently set
equal to
the identity $I$.

A quantum determinant can be defined, and is central
 only in a subclass of the multiparametric
deformations. In this subclass, however, the element $u$ is not
central.
\sk
In Section 6 we project the
bicovariant differential calculus of $GL_{q,r}(N+1)$
to $IGL_{q,r}(N)$ and
study the bicovariant bimodules of $1$-forms and
tangent vectors on
$IGL_{q,r}(N)$. In particular, the $q$-Lie algebra is given
explicitly. We also study in detail
the exterior algebra and the exterior derivative, and  find the
Cartan-Maurer equations.
\sk
In Section 7 we discuss the multiparametric quantum plane,
i.e. the quantum coset space  $\IGLqrN / \GLqrN$
spanned by the
coordinates $x^a$, and find a generalization
of the differential geometry of the $q$-plane
of \cite{qplane} (see also Schirrmacher in \cite{multiparam}).
\sk
In the Table at the
end of the paper we specialize
our general treatment to  $\IGLqrtwo$
and collect
all the relevant  formulas for  its
bicovariant differential calculus.
\sk

\sect{$GL_{q,r}(N)$ and its real forms}

First, we recall the definition of $GL_{q,r}(N)$.
It is the algebra
(over the complex field)
freely generated by the
non-commuting matrix elements $\T{A}{B}$,
({\small \sl  A,B=1,..N}),
 the identity $I$ and  the inverse $\Xi$ of
the $q$-determinant of $T$ defined in (\ref{qrdet}), modulo
the ``$RTT$" relations:
\eq
\R{AB}{EF} \T{E}{C} \T{F}{D} = \T{B}{F} \T{A}{E} \R{EF}{CD}
\label{RTTGL}
\en
\noi where the $R$-matrix is given by \cite{Schirrmacher}:
\eq
\R{AB}{CD}=\de^A_C \de^B_D [{r\over q_{AB}}+(r-1) \de^{AB}]
+(r-r^{-1})~ \de^A_D
\de^B_C
 \theta^{AB} \label{Rmp}
\en
\noi with $\theta^{AB}=1$
for {\footnotesize $A > B$ } and zero otherwise, and
\eq
q_{AB}={r^2\over q_{BA}},~~q_{AA}=r  \label{qabqba}
\en

It is useful to list the nonzero
complex
components of the $R$ matrix (no sum on repeated indices):
\eqa
& &\R{AA}{AA}=r \cr
& &\R{AB}{AB}={r \over q_{AB}} , ~~~~~~~\mbox{\footnotesize
 $A \not= B$ }\cr
& &\R{BA}{AB}=r-r^{-1},~~~\mbox{\footnotesize $B>A $}
\label{Rnonzero}
\ena

The $R$ matrix in (\ref{Rmp}) satisfies the
quantum Yang-Baxter (QYB)
equation:
\eq
\R{A_1B_1}{A_2B_2} \R{A_2C_1}{A_3C_2} \R{B_2C_2}{B_3C_3}=
\R{B_1C_1}{B_2C_2} \R{A_1C_2}{A_2C_3} \R{A_2B_2}{A_3B_3}.
\label{YB}
\en

The standard uniparametric $R$ matrix \cite{FRT}
is obtained from
(\ref{Rmp}) by setting all deformation
parameters $q_{AB},r$ equal to a
single parameter $q$.
\sk
The quantum determinant of $T$ and
its inverse $\Xi$ are defined by:
\eq
\Xi~ {\det} T={\det} T~\Xi=I\label{Xi}
\en
\eq
{\det} T \equiv \sum_{\sigma}  \left [
\prod_{A<B,\sigma (A) > \sigma(B)}
 (-{r^2\over q_{\sigma(B) \sigma(A)}})  \right ]~
\T{1}{\sigma(1)} \cdots
\T{N}{\sigma(N)} \label{qrdet}
\en

{\sl Note 1:} In the uniparametric case $r=q_{AB}=q$ we recover
 the usual formula
\eq
\detq T \equiv \sum_{\sigma} (-q)^{l(\sigma)}
\T{1}{\sigma(1)} \cdots
\T{N}{\sigma(N)} \label{qdet}
\en
\noi where $l(\sigma)$ is the minimum number of
transpositions in the permutation
$\sigma$.
\sk
{\sl Note 2:}
In more mathematical terms, the algebra $GL_{q,r}(N)$ is the quotient
of the non-commuting algebra ${\bf C}\langle T^A{}_B, I, \Xi\rangle$
freely
generated by the elements $T^A{}_B, I, \Xi$ with respect to
the two-sided ideal in
${\bf C}\langle T^A{}_B, I, \Xi\rangle$
generated by the $RTT$ relations (\ref{RTTGL}).
\sk
{\sl Note 3:} the
inverse matrix $R^{-1}$, defined as
\eq
\Rinv{AB}{CD}\R{CD}{EF} \equiv \de^
A_E \de^B_F \equiv \R{AB}{CD}\Rinv{CD}{EF}.
\en
\noi is given by
\eq
R^{-1}_{q,r}=R_{\qm,\rminus} \label{Rinv}
\en
\sk
{\sl Note 4:} the $\Rh$ matrix defined by
$\Rhat{AB}{CD} \equiv \R{BA}{CD}$
satisfies the spectral decomposition
(Hecke condition):
\eq
(\Rh-rI)~ (\Rh + r^{-1} I)=0 \label{Hecke}
\en
\sk
 {\sl Note 5:} the determinant in (\ref{qrdet}) is
central if and only if the following conditions on the
 parameters are satisfied
(see ref. \cite{Schirrmacher}):
\eq
q_{1,A} q_{2,A} \cdots q_{A-1,A} {r^2\over q_{A,A+1}}
{r^2\over
q_{A,A+2}}
\cdots {r^2\over q_{A,N}}=const. \label{centralitycondition}
\en
\noi for all {\small A=1,...N}. This results in
{\small N-1} conditions among
the $q_{AB}$ and determines $const=r^{N-1}$. Using
 (\ref{qabqba}), and defining
\eq
Q_A \equiv
\prod_{C=1}^N ({q_{CA}\over r})
\label{Qdefinition}
\en
the centrality conditions
(\ref{centralitycondition}) become:
\eq
Q_A=1 \label{centralityconditionQ}
\en
We have used also $const=r^{N-1}$, so that
only $N-1$ of the above conditions are independent. Indeed
the $Q_A$ satisfy the relation
\eq
Q_1 Q_2 \cdots Q_N=1
\en
In  general we have:
\eq
(\detqr T ) \T{A}{B} = {Q_A \over Q_B} \T{A}{B} (\detqr T),
{}~~\Xi \T{A}{B} = {Q_B \over Q_A} \T{A}{B} \Xi
\label{detcomm}
\en
When (\ref{centralityconditionQ})
holds, we can consistently set $\detqr \T{A}{B} = I = \Xi$,
 and obtain
the multiparametric deformations $SL_{q,r}(N)$.
\sk
The  $RTT$ equation (\ref{RTTGL}) and the
quantum Yang--Baxter equation can be rewritten
more compactly as:
\eq
R_{12} T_1 T_2 = T_2 T_1 R_{12} \label{rtt}
\en
\eq
R_{12} R_{13} R_{23}=R_{23} R_{13} R_{12}, \label{yb}
\en
\noi where the subscripts 1, 2 and 3 refer to
different couples of
indices. Thus $T_1$ indicates the matrix
$\T{A}{B}$, $T_1 T_1$
indicates $\T{A}{C} \T{C}{B}$, $R_{12} T_2$
indicates $\R{AB}{CD} \T{D}{E}$
and so on, repeated subscripts meaning matrix
multiplication.

\sk
The algebra $GL_{q,r}(N)$ becomes a Hopf algebra with
the following coproduct
$\D$, counit $\epsi$ and coinverse $\kappa$:
\eqa
& & \D(\T{A}{B})=\T{A}{B} \otimes \T{B}{C}  \label{cos1} \\
& & \epsi (\T{A}{B})=\delta^A_B\\
& & \kappa(\T{A}{B})=\Ti{A}{B} \label{coinverse}\\
& &  \D (\detqr T)=\detqr T \otimes \detqr T, ~~
\D (\Xi)=\Xi \otimes
\Xi,
{}~~\D(I)=I\otimes I\\
& &  \epsi (\detqr T)=1,~~\epsi (\Xi)=1,~~\epsi (I)=1\\
& &  \kappa (\detqr T)=\Xi,~~\kappa (\Xi)=
\detqr T,~~\kappa (I)=I
\label{cos2}
\ena
\noi The quantum inverse of $\T{A}{B}$ in
(\ref{coinverse}) is given
by:
\eq
\Ti{A}{B}=\Xi~\pp{AB}{1,N}~ t_B^{~A}  \label{Tinverse}
\en
\noi where $t_B^{~A}$ is the quantum minor, i.e. the quantum
determinant
of the submatrix of $T$ obtained by removing the
{\small $B$}-th row
and
the {\small $A$}-th column, and $\pp{AB}{1,N}$ is a
function of the
parameters $q$:
\eq
\pp{AB}{1,N} \equiv  { {\prod_{C=B+1}^N (-q_{BC})}  \over
 {\prod_{D=A+1}^N (-q_{AD})}} \label{defPi}
\en
\noi The superscript (1,N) reminds the range
of the indices {\small A,B,C,..}.
 In the uniparametric case, the quantum
inverse
has the simpler expression:
\eq
\Ti{A}{B}=\Xi~ (-q)^{A-B} t_B^{~A}
\en
{\sl Note 5:} In general $\kappa^2 \not= 1$. The following
useful relation holds
\eq
\kappa^2 (\T{A}{B})=\Dmat{A}{C} \T{C}{D}
 \Dmatinv{D}{B}=d^A d^{-1}_B
\T{A}{B},  \label{k2}
\en
\noi where $D$ is a diagonal matrix,
$\Dmat{A}{B}=d^A \de^A_{B}$, given
by $d^A=r^{2A-1}$ for $\GLqrN$. This matrix satisfies:
\eq
d^A d^{-1}_C \Rinv{BA}{DC} \R{EC}{BF}=
\de^A_F \de^E_D,~~~
d^A d^{-1}_C \R{AB}{CD} \Rinv{CE}{FB}=
\de^A_F \de^E_D  \label{Rsecinv1}
\en
\eq
d^B d^{-1}_D \Rinv{AB}{CD} \R{CE}{FB}=
\de^A_F \de^E_D,~~~
d^B d^{-1}_D \R{BA}{DC} \Rinv{EC}{BF}=
\de^A_F \de^E_D \label{Rsecinv2}
\en
\eq
\R{AC}{CB} d^{-1}_C=\de^A_B=\Rinv{AC}{CB} d_{C}
\label{RD}
\en
\noi Relations (\ref{Rsecinv1}) and (\ref{Rsecinv2})
define a second inverse
$R^{\sim 1}$ of
the $R$ matrix  and a second inverse $(R^{-1})^{\sim 1}$
of the $R^{-1}$ matrix as:
\eq
\Rsecondinv{AB}{CD} \equiv
d^B d^{-1}_D \Rinv{AB}{CD}
\en
\eq
\Rinvsecondinv{AB}{CD} \equiv
d^A d^{-1}_C \R{AB}{CD}
\en
\noi Using (\ref{RD})
we can relate the $D$ matrix to this second inverse:
 \eq
\Dmatinv{A}{B}=\Rsecondinv{AC}{CB},~~~\Dmat{A}{B}=
\Rinvsecondinv{AC}{CB}
\en
This generalizes the analogous discussion for
the uniparametric $D$ matrix
given in \cite{FRT}.

\sk
We turn now to the real forms of $\GLqrN$.
These are defined by
*-involutions of the $\GLqrN$ Hopf algebra, that is
mappings which are algebra antimorphisms and a co-algebra
automorphisms:
\eq
(\lambda  a)^*= {\bar \lambda} a^*,~~~(ab)^*=b^* a^*,
{}~~~\D (a^*)=[\D (a)]^*~~~\lambda \in {\bf C};~~a,b \in \GLqrN
\label{hopfinvo1}
\en
\noi (${\bar \lambda}$ is the usual complex
conjugate of $\lambda$)
and satisfy the involution conditions:
\eq
(a^*)^*=a,~~~\kappa ( [\kappa (a^*)]^*)=a~~~\forall
a \in  \GLqrN
\label{hopfinvo2}
\en
Moreover, these involutions (also called conjugations)
must be compatible with the $RTT$ relations:
this restricts the range of the parameters $q,r$.
Three such conjugations are known (cf. \cite{Schirrmacher}):
\sk
i)   $~~T^*=T$, i.e. the elements $\T{A}{B}$ are ``real".
Applying
the *-conjugation to the $RTT$ equations (\ref{RTTGL})
yields again the $RTT$ relations if the $R$ matrix satisfies
${\bar R}=R^{-1}$. This happens for
$|q_{AB}|=|r|=1$, i.e. for
deformation parameters lying on the unit circle in ${\bf C}$
(cf. eq. (\ref{Rinv})).  The quantum group is then
denoted by
$GL_{q,r}(N;{\bf R})$.
\sk
ii)   $~~(\T{A}{B})^*=\T{A'}{B'}$ with primed indices
defined as {\small $A'=N+1-A$}.  Here compatibility
with the $RTT$ relations requires
$\Rbar{AB}{CD}=\R{B'A'}{D'C'}$,
satisfied when ${\bar q}_{AB}=q_{B'A'},~r \in {\bf R}$.
We can then define ``real" generators as $(T+T^*)/2$,
and the corresponding quantum group could be called
$GL^\prime_{q,r}(N;{\bf R})$.
\sk
iii) $~~(\T{A}{B})^*=\kappa (\T{B}{A})$, the
generalization of the
unitarity
condition for the matrix $T$. In this case (left as
an exercise in
\cite{Schirrmacher}) the restriction on the $R$
matrix is $\Rbar{AB}{CD}=\R{DC}{BA}$, leading to
the conditions
${\bar q}_{AB}=q_{BA},~r \in {\bf R}$. The
corresponding quantum
groups are denoted by $U_{q,r}(N)$.
\sk
Imposing  also $\detqr T=I$ yields the quantum groups
$SL_{q,r}(N;{\bf R})$,  $SL^{\prime}_{q,r}(N;{\bf R})$ and
$SU_{q,r}(N)$ respectively.

%%%%%%%%%%%%%%%%%%%%%%%%%%%%%%%%%

\sect{The universal enveloping algebra of $\GLqrN$}

%%%%%%%%%%%%%%%%%%%%%%%%%%%%%%%%%

We construct the universal enveloping algebra of $\GLqrN$ as the
algebra of regular functionals \cite{FRT} on $\GLqrN$:
it is generated
by the functionals $\LLpm , \epsi$ and $\Phi$ defined below.

\sk
\noi {\bf Algebra structure}
\sk

The $\LLpm$ linear functionals  on $GL_{q,r}(N)$ are defined
by their value on the matrix elements $\T{A}{B}$  :
\eq
\Lpm{A}{B} (\T{C}{D})= \Rpm{AC}{BD}, \label{LonT}
\en
\eq
\Lpm{A}{B} (I)=\de^A_B \label{LonI}
\en
\noi with
\eq
\Rp{AC}{BD} \equiv c^+ \R{CA}{DB} \label{Rplus}
\en
\eq
\Rm{AC}{BD} \equiv c^- \Rinv{AC}{BD}, \label{Rminus}
\en
\noi where $c^+$, $c^-$ are free parameters (see later).
\sk
To extend the definition (\ref{LonT})
to the whole algebra $GL_{q,r}(N)$ we set
\eq
\Lpm{A}{B} (ab)=\Lpm{A}{C} (a) \Lpm{C}{B} (b),
{}~~~\forall a,b\in
GL_{q,r}(N)
\label{Lab}
\en
\sk
The commutations between $\Lpm{A}{B}$
and $\Lpm{C}{D}$ are induced by those between the $T$'s:
\eq
R_{12} \LLpm_2 \LLpm_1=\LLpm_1 \LLpm_2 R_{12} \label{RLL}
\en
\eq
R_{12} \LLp_2 \LLm_1=\LLm_1 \LLp_2 R_{12}, \label{RLpLm}
\en
\noi where as usual the product $\LLpm_2 \LLpm_1$
is the convolution
product $\LLpm_2 \LLpm_1 \equiv (\LLpm_2 \otimes \LLpm_1)\D$.
\sk
{\sl Note 1 :} $L^+$ is upper
triangular and $L^-$ is lower triangular. Proof:
apply $L^+ $ and
$L^-$ to
 the $T$ elements
and use the upper and lower triangularity of $R^+$ and $R^-$,
respectively.
\sk
A determinant can be defined for the matrix
$\Lpm{A}{B}$ as in (\ref{qrdet})
with $q \rightarrow \qm,~r \rightarrow \rminus$.
Indeed the ``$RLL$"
relations are identical to the  "$RTT$" with
$R \rightarrow R^{-1}$ (which means
 $q \rightarrow \qm,~r \rightarrow \rminus$,
cf. eq. (\ref{Rinv})). Then , because
of the upper or lower triangularity of $\LLp$ and $\LLm$
respectively, we have
\eq
\detqrm \LLpm=\Lpm{1}{1} \Lpm{2}{2} \cdots \Lpm{N}{N} \label{detL}
\en

\noi A quantum inverse for $\Lpm{A}{B}$ can be found, using
an expression analogous to (\ref{Tinverse})
with $q_{AB} \rightarrow
\qm_{AB}$.For this we need to introduce the
element $\Phi$ defined by:
\eq
\Phi\detqrm L^+\detqrm L^- =  \detqrm L^+\detqrm L^- \Phi = \epsi ~.
\en
Then the quantum inverse of
 $\Lpm{A}{B}$ is given by:
\eq
(\Lpm{A}{B})^{-1}=\Phi ~\detqrm L^{\mp} ~ \pp{BA}{1,N}
{}~ \ell_B^{~A}  \label{Linverse}
\en
\noi where $\ell_B^{~A}$ is the quantum minor and $\pp{BA}{1,N}$
is given in (\ref{defPi}). Notice that $\Phi ~\detqrm L^{\mp}$ is the
inverse of $\detqrm L^{\pm}$ because of property (\ref{LL3}) below.
\sk
\noi {\bf Coalgebra structure}
\sk
The co-structures of the algebra generated by the
functionals $L^{\pm}$, $\epsi$ and $\Phi$
are defined by the duality (\ref{LonT}):
%%%%%%%%
%\eq
%\Dp(\Lpm{A}{B})(\T{C}{D} \otimes \T{E}{F}) \equiv \Lpm{A}{B}
%(\T{C}{D}\T{E}{F})=\Lpm{A}{G}(\T{C}{D}) \Lpm{G}{B} (\T{E}{F})
%\en
%\eq
%\ep (\Lpm{A}{B})\equiv \Lpm{A}{B} (I)
%\en
%\eq
%\kp (\Lpm{A}{B})(\T{C}{D})\equiv \Lpm{A}{B} (\kappa (\T{C}{D}))
%\en
\eq
\Dp(\Lpm{A}{B})(a \otimes b) \equiv \Lpm{A}{B}
(ab)=\Lpm{A}{G}(a) \Lpm{G}{B} (b)
\en
\eq
\ep (\Lpm{A}{B})\equiv \Lpm{A}{B} (I)
\en
\eq
\kp (\Lpm{A}{B})(a)\equiv \Lpm{A}{B} (\kappa (a))
\en
\noi so that
\eqa
& & \Dp (\Lpm{A}{B})=\Lpm{A}{G} \otimes \Lpm{G}{B}\label{copLpm}\\
& & \ep (\Lpm{A}{B})=\de^A_B \label{couLpm}\\
& & \kp (\Lpm{A}{B})=\Lpm{A}{B} \circ \kappa \label{coiLpm}\\
& &\Dp(\detqrm \LLpm)=\detqrm \LLpm \otimes \detqrm\LLpm,\\
 & &\Dp(\Phi)=\Phi\otimes\Phi, \Dp(\epsi)=\epsi\otimes \epsi\\
& &\ep(\detqrm\LLpm)=1,~\ep(\Phi)=1,~\ep(\epsi)=1\\
& &\kp(\detqrm\LLpm)=\Phi~\detqrm\LLmp,~\\
& &\kp(\Phi)=\detqrm \LLp \detqrm \LLm,~\kp(\epsi)=\epsi
\ena
\sk
\noi {\sl Note 2 :} In (\ref{coiLpm}) we have defined
$\kp$ using $\kappa$,
we now prove that $\kp(\Lpm{A}{B})=(\Lpm{A}{B})^{-1}$
as defined in (\ref{Linverse}).
 This shows that \(\kp(\Lpm{A}{B})\) is expressible by
polynomials in
\(\Lpm{A}{B}, \Phi\).
\sk
\noi {\sl Proof :} From  \((\LLpm)^{-1}\LLpm=\epsi\) we have
\({ 1}=[(\LLpm_1)^{-1}\LLpm_1](T)=(\LLpm_1)^{-1}(T_2)\LLpm_1(T_2)=
(\LLpm_1)^{-1}(T_2)R_{12}^{\pm}\) so that
\((\LLpm_1)^{-1}(T_2)={R_{12}^{\pm}}^{-1}\).

\noi {}From $\kappa(T)T= 1$
we similarly have
\([\kp(\LLpm_1)](T_2)={R_{12}^{\pm}}^{-1}\) and therefore
$\kp(\Lpm{A}{B})=(\Lpm{A}{B})^{-1}$.
\sk

Since $\kp$ is an inner operation in the algebra generated by the
functionals \(\Lpm{A}{B},~\epsi\) and \(\phi\) we
conclude that these
elements generate the Hopf algebra $\UglqrN$  of the regular
functionals on the quantum
group $\GLqrN$.
\sk

In the following we list some useful properties of the $\LLpm$
functionals.
\sk
\noi {\bf Properties of $\LLpm$}
\sk
\noi {\bf i)} From (\ref{RLL}) and (\ref{RLpLm}) we have
\eq
\Lpm{A}{A} \Lpm{B}{B}=\Lpm{B}{B} \Lpm{A}{A}~;~~
\Lp{A}{A} \Lm{B}{B}=\Lm{B}{B} \Lp{A}{A} \label{LL2}
\en
\noi As a consequence:
\eq
\detqrm \LLp \detqrm \LLm = \detqrm \LLm \detqrm \LLp.
\label{LL3}
\en

\noi {\bf ii)} From (\ref{LonT}) we deduce:
\eq
\Lpm{A}{B} (\detqr T)= \de^A_B (c^{\pm})^N r^{\pm 1}
Q_A^{-1}\label{LondetT}
\en
{\sl Proof:} observe that $\Lpm{A}{B}
(\detqr T)=\Lpm{A}{B} (\T{1}{1}
\T{2}{2}
\cdots \T{N}{N})$ since all the other permutations
do not contribute,
due to the structure of the $R^{\pm}$ matrix. Then it
is easy to see that
\eq
\Lpm{A}{B} (\T{1}{1} \T{2}{2} \cdots \T{N}{N})=
\nonumber
\en
\eq
\de^A_B (c^{\pm})^N
\Rpm{A1}{A1} \Rpm{A2}{A2} \cdots \Rpm{AN}{AN}=
\de^A_B (c^{\pm})^N
r^{\pm 1} Q_A^{-1}
\en
As a corollary, we have that :
\eq
\Lpm{A}{B} (\Xi)=\de^A_B (c^{\pm})^{-N}
r^{\mp 1} Q_A \label{LonXi}
\en
\noi  {\sl Proof}: use (\ref{Xi}) and (\ref{Lab}) .
\sk

\noi {\bf iii)} From the expression
(\ref{LL4}) of $\detqrm \LLpm$ we have
\eq
\detqrm \LLpm (\T{A}{B})=\de^A_B (c^{\pm})^N
 r^{\pm 1}Q_A \label{LL4}
\en
%\eq
%\detqrm \LLpm (I)=1 \label{LL5}
%\en
\noi {\sl Proof:}
\eq
\detqrm \LLpm (\T{A}{B})=\de^A_B (c^{\pm})^N
\Rpm{1A}{1A} \cdots
\Rpm{NA}{NA}=
\de^A_B (c^{\pm})^N r^{\pm 1} Q_A.
\en
As a corollary we obtain
\eq
\detqrm \LLpm (\detqr T) = \det R^{\pm} =
(c^{\pm})^{N^2} r^{\pm N}
\label{LL6}
\en
\noi where the notation det$R^{\pm}$ means the
ordinary determinant
of the square matrix $\Rpm{AB}{CD}$ where rows
and columns are
respectively
labelled by the combined indices AB and CD.
Notice that
\eq
\detqrm \LLpm (\T{A}{B})= \Lpm{A}{B} (\detqr T) Q_A^2.
\en
\noi {}From (\ref{detL}) it is easy to see that
$\detqrm \LLpm (I)=1 \label{LL5}.$
\sk
\noi {\bf iv)}  Since the $RLL$ relations are
the same as the $RTT$ relations with
$q_{AB}\rightarrow(q_{AB})^{-1}$,  $r \rightarrow r^{-1}$,
we obtain a formula analogous to (\ref{detcomm}):
\eq
(\detqrm \LLpm)\Lpm{A}{B}={Q_B\over Q_A}\Lpm{A}{B}
(\detqrm \LLpm)
\label{LpmdetqrmLpm}.
\en
We also have
\eq
(\detqrm \LLmp)\Lpm{A}{B}={Q_B\over Q_A}
\Lpm{A}{B}(\detqrm \LLmp)
\label{LpmdetqrmLmp}
\en
\sk
\noi {\bf v)} From (\ref{LpmdetqrmLpm}) and
(\ref{LpmdetqrmLmp}) the following
element:
\eq
\detqrm \LLp (\detqrm \LLm)^{-1}=
(\detqrm \LLm)^{-1}\detqrm \LLp
\en
is seen to be central.  Notice that it is
also group-like since
\eq
\Dp (\detqrm L^{\pm})=\detqrm L^{\pm} \otimes
\detqrm L^{\pm}.
\en
In general even if $\detqrm \LLp (\detqrm \LLm)^{-1}$
is central and group-like it is not equal to $\epsi$ because
\eq
\detqrm \LLp (\detqrm \LLm)^{-1}(\T{A}{B})=
(c^+)^N (c^-)^{-N} ~r^{2}\delta^{A}_{B}.
\en
\sk
\noi {\bf vi)} The elements $\Lp{A}{A} \Lm{A}{A}$
(no sum on {\small $A$})
 play a special role for particular
values of the deformation parameters $q_{AB}, r$;  if we set
\eq
\Lp{A}{A} \Lm{A}{A} \equiv \epsi_A
\en
\noi we leave as an exercise to deduce
that (no sum on repeated indices):
\eq
\epsi_A(\T{B}{C})\equiv c^+ c^- \de^B_C ~{q^2_{AB} \over r^2},
{}~~\epsi_A(I)=1, ~~
\epsi_A(\Xi)=[\epsi_A (\detqr T)]^{-1}
\en
\eq
\epsi_A (ab)=\epsi_A (a) \epsi_A (b), ~~~\mbox{$a,b ~\in \GLqrN$}
\en
\eq
\kp (\Lpm{A}{A})=\Lmp{A}{A}\epsi_A^{-1}
\en
\eq
\epsi_A \epsi_B=\epsi_B \epsi_A,
{}~~~~\epsi_A \Lpm{B}{B}=\Lpm{B}{B} \epsi_A
\en
\eq
\detqrm \LLp \detqrm \LLm =
\epsi_1 \cdots \epsi_N~;
\en
\eq
\kp (\detqrm L^{\pm})=\detqrm L^{\mp}
(\epsi_1 \cdots \epsi_N)^{-1}=(\epsi_1 \cdots \epsi_N)^{-1}
\detqrm L^{\mp}
\en
\sk

\noi {\sl Note 3}: When $\detqr T$ is central
($Q_A=1$) we also have that
$\det \LLpm$ is central (cf.  (\ref{LpmdetqrmLpm})
and (\ref{LpmdetqrmLmp}) ).
If we set  $\detqr T =I$,
then $\Lpm{A}{B} (\detqr T)= \de^A_B (c^{\pm})^N r^{\pm 1}  $
must be equal to $\de^A_B$,
or $c^{\pm}=r^{\mp {1\over N}} \al^{\pm}$
with $(\al^{\pm})^{N}=1$.
In this case $[\detqrm \LLpm](\T{A}{B})=\delta^A_B$
so that $\det \LLpm=\epsi$. Thus for $Q_A=1,
 (c^{\pm})^{N} r^{\pm 1}=1 $, the functionals $\LLpm$and
$\epsi$ generate
the Hopf algebra $U(sl_{q,r}(N))$, and we have
the simplified relations:
\eqa
& & \detqrm \LLp (\detqrm \LLm)^{-1}=\epsi\\
& & [\Lpm{A}{B}](\detqr T)=\delta^A_B~~
{\mbox{ no sum on {\small$A$}}}\\
& & [\detqrm \LLpm](\T{A}{B})=\delta^A_B\\
& & [\detqrm \LLpm](\detqr T)=1
\ena
\sk
\noi {\sl Note 4:} When $q_{AB}=r $  we recover
the standard uniparametric $R$ matrix.
We have also  $Q_A=1$ and
\eq
\forall\, A ~~\epsi_A=\epsi
{}~~\mbox{ i.e. }
\Lp{A}{A}\Lm{A}{A}=\Lm{A}{A}\Lp{A}{A}
=\epsi \label{epsiaepsi}
\en
In this case the Hopf algebra of functionals
$\UglqrN$ is equivalent to
the algebra generated by  the symbols $\LLpm , \Phi$
and $\epsi$
modulo relations
(\ref{RLL}),(\ref{RLpLm}) and
(\ref{epsiaepsi}).
\sk
\noi {\sl Note 5:}  $\GLqrN$ and $U(gl_{q,r}(N))$
are graded Hopf algebras:
$T^A{}_B$ has grade $+1$,
$\kappa(T^A{}_B)$ has grade $-1$,
$I$ has grade $0$,
$\det T$ has grade $+N$ etc., and similarly for $L^{\pm}$.
\sk

\noi {\bf Conjugation}
\sk

The *-conjugation on $\GLqrN$ induces a *-conjugation
on
$\UglqrN$ in two possible ways
(we denote them as $*$ and $\sharp$-conjugations):
\eq
L^* \equiv {\overline {L(\kappa  (a^*))}}
\en
\eq
L^{\sharp} \equiv {\overline {L(\kappa^{-1} (a^*))}}
\en
\noi the overline being the usual complex conjugation.
Both $*$ and $\sharp$ can be shown
to satisfy all the properties of Hopf algebra involutions
(\ref{hopfinvo1}), (\ref{hopfinvo2}).  It is not difficult
to determine their action
on the basis elements $\Lpm{A}{B}$.
The three $\GLqrN$ $*$-conjugations  i), ii), iii) of
the previous section induce
respectively the following conjugations on the
$\Lpm{A}{B}$:
\eqa
& &i) ~~~~(\Lpm{A}{B})^*=\Lpm{A}{B} \cr
& &ii) ~~~(\Lpm{A}{B})^*=\Lmp{A'}{B'} \cr
& &iii)~~(\Lpm{A}{B})^*=\kpm~  (\Lmp{B}{A})
\ena
\eqa
& &i) ~~~~(\Lpm{A}{B})^{\sharp}=\kpsq~ (\Lpm{A}{B}) \cr
& &ii) ~~~(\Lpm{A}{B})^{\sharp}=\kpsq~ (\Lmp{A'}{B'}) \cr
& &iii)~~(\Lpm{A}{B})^{\sharp}=\kp~ (\Lmp{B}{A})
\ena
\noi so that $(\Lpm{A}{B})^{\sharp}=\kpsq~[ (\Lpm{A}{B})^*]$.

%%%%%%%%%%%%%%%%%%%%%%%%%%%%%%%%%

\sect{Differential calculus on $\GLqrN$}

%%%%%%%%%%%%%%%%%%%%%%%%%%%%%%%%%

The bicovariant differential calculus on the uniparametric
$q$-groups of the $A,B,C,D$ series can be formulated in terms
of the corresponding $R$-matrix, or equivalently in terms of
the $\LLpm$ functionals.  This holds also for the multiparametric
case.  In fact all formulas are the same, modulo substituting
the $q$ parameter with $r$  when
it appears explicitly (typically as ${1 \over {q-q^{-1}}}$).
\sk
We briefly recall how to construct a bicovariant
calculus.  The general procedure can be found in ref.
\cite{Jurco}, or, in the notations we adopt
here, in ref. \cite{Aschieri1}.
It realizes the axiomatic construction of ref. \cite{Wor}.
\sk

\noi { \bf The space of quantum 1-forms}
\sk
As in the uniparametric case \cite{Jurco}, the functionals
\eq
\ff{A_1}{A_2B_1}{B_2} \equiv \kp (\Lp{B_1}{A_1}) \Lm{A_2}{B_2}.
\label{defff}
\en
and the elements of $A=\GLqrN$:
\eq
\MM{B_1}{B_2A_1}{A_2} \equiv \T{B_1}{A_1} \kappa (\T{A_2}{B_2}).
\label{defMM}
\en
satisfy the following relations, where for simplicity
we use the adjoint indices $ i,j,k,...$
with ${}^i={}_A^{~B},~{}_i={}^A_{~B}~:$
\eqa
& & \f{i}{j} (ab)= \f{i}{k} (a) \f{k}{j} (b) \label{propf1}\\
& & \f{i}{j} (I) = \del^i_j ~\label{propf2} \\
& & \Delta (\M{j}{i}) = \M{j}{l} \otimes \M{l}{i} \label{copM}\\
& & \epsi (\M{j}{i}) = \delta^i_j \label{couM} \\
& & \M{i}{j} (a * \f{i}{k})=(\f{j}{i} * a) \M{k}{i} \label{propM}
\ena
\noi The star product between a functional on $A$ and an
element of $A$ is defined as:
\eqa
& & \chi * a \equiv (id \otimes \chi) \D (a)  \\
& & a * \chi \equiv (\chi \otimes id) \D (a),~~~~~
a \in A, ~\chi \in A'
\ena
Relation  (\ref{propM}) is easily checked for
$a=\T{A}{B}$ since in
this case it is implied by the $RTT$ relations;
it holds for a generic $a$ because of property
(\ref{propf1}).
\sk
The space of quantum one-forms is defined as a right
$A$-module $\Ga$
freely generated  by the  {\sl symbols} $\ome{A_1}{A_2}$:
\eq
\Ga\equiv a^{A_1}_{~A_2}\ome{A_1}{A_2} ,~~~a^{A_1}_{~A_2} \in A
\en
{\sl Theorem} (due to Woronowicz: see Theorem 2.5  in
the last ref. of \cite{Wor},
p. 143): because of the
properties  (\ref{propf1})-(\ref{propM}), $\Ga$
becomes a bimodule over $A$ with the following
right multiplication:
\eq
\ome{A_1}{A_2} a=(\ff{A_1}{A_2B_1}{B_2} * a)~\ome{B_1}{B_2},
\label{omea}
\en
In particular:
\eq
\ome{A_1}{A_2} \T{R}{S}=s  \Rinv{TB_1}{CA_1}
\Rinv{A_2C}{B_2S} \T{R}{T}\ome{B_1}{B_2} \label{commomT}
\en
\eq
\ome{A_1}{A_2} \det T=s^N r^{-2} {Q_{A_1} \over Q_{A_2}}
 (\det T) \ome{A_1}{A_2} \label{commomdetT}
\en
\eq
\ome{A_1}{A_2} \Xi=s^{-N} r^{2} {Q_{A_2} \over Q_{A_1}}
 (\Xi) \ome{A_1}{A_2} \label{commomXi}
\en
Moreover we can  define  a left and a right action of
$\GLqrN$ on $\Ga$:
\eqa
& & \DL:\Ga\rightarrow A\otimes
\Ga~~;~~~\DL(a^{A_1}{}_{A_2}\ome{A_1}{A_2})
\equiv \D(a^{A_1}{}_{A_2})I\otimes \ome{A_1}{A_2}
\label{deltaLomega} \\
& & \DR:\Ga\rightarrow  \Ga\otimes A
{}~~;~~~\DR(a^{A_1}{}_{A_2}\ome{A_1}{A_2})
\equiv \D(a^{A_1}{}_{A_2})\ome{B_1}{B_2}\otimes
\MM{B_1}{B_2A_1}{A_2}.
\label{deltaRomega}
\ena
These actions commute
\eq
(id \otimes \DR) \DL = (\DL \otimes id) \DR
\en
and give a bicovariant bimodule structure to $\Ga$.
\sk
{\sl Note: }: $\det T$ and $\Xi$ commute with all the $\om$
(and thus can be set to $I$) iff
all $Q_A$ are equal and for $s^N r^{-2}=1$, or
$s=r^{{2\over N}} \al$ with $\al^N =1$,
which agrees with  the condition found in Note 3 of previous section.
\sk

\noi {\bf Exterior derivative}
\sk
\noi   A derivative operator
$d ~:~ A\longrightarrow \Ga$ can be defined via the element
$\tau\equiv \sum_A \ome{A}{A} \in \Ga$. This element
is
easily shown to be
left and right-invariant:
\eq
\DL(\tau)=I\otimes\tau~~;~~~
\DR(\tau)=\tau\otimes I
\en
and defines the derivative  $d$  by
\eq
da=\rinv [\tau a - a \tau]. \label{defd1}
\en
\noi The factor $\rinv$ is necessary for a correct
classical limit $\rone$.
It is immediate to prove the Leibniz rule
\eq
d(ab)=(da)b+a(db),~~\forall a,b\in A ~. \label{Leibniz}
\en
Another expression for the derivative is given by:
\eq
da = ( \cchi{A_1}{A_2} * a) ~\ome{A_1}{A_2} \label{defd2}
\en
where
\eq
\cchi{A}{B} = \rinv [\ff{C}{CA}{B}-\de^A_B \epsi]
\label{defchi}
\en
are the left-invariant vectors dual to the left-invariant
1-forms $\ome{A_1}{A_2}$.
The equivalence of  (\ref{defd1}) and (\ref{defd2}) can be
shown by using the rule (\ref{omea}) for $\tau a$ in
the right-hand side of (\ref{defd1}).

Using (\ref{defd2}) we compute the exterior derivative
on the basis elements of $\GLqrN$, and on the $q$-determinant:
\eq
d ~\T{A}{B}=\rinv [s~\Rinv{CR}{ET} \Rinv{TE}{SB}
\T{A}{C}-\de^S_R \T{A}{B}]
{}~\ome{R}{S} \label{dTAB}
\en
\eq
d~ \Xi={{s^{-N} r^2-1} \over {r-\rminus}} ~\Xi~ \tau \label{dXi}
\en
\eq
d~ \detqr T={{s^N r^{-2}-1}\over{r-\rminus}}
{}~(\detqr T) \tau \label{ddetT}
\en
The reader can verify via the Leibniz rule,  and with the
help of eq. (\ref{commomdetT}),
that $d [(\det T)\Xi]=d [\Xi (\det T)]=0$.

\noi {\sl Note: } again $\det T=I=\Xi$ requires
$s^N r^{-2}=1$.
\sk
\noi
Every element  $\rho$ of $\Gamma$, which by definition is
written in a unique way as
$\rho=a^{A_1}{}_{A_2}\ome{A_1}{A_2}$,  can also be written as
\eq
\rho=a_k db_k  \label{propd}
\en
\noi for some $a_k,b_k$ belonging to $A$. This can
be proven directly by inverting
the relations (\ref{dTAB}) and (\ref{dXi}), after
replacing the explicit
values of the $R^{-1}$ matrices. The
result  is an expression of
 the $\om$ in terms of a linear
combination of $\kappa (T) dT$, as in the classical case:
\eqa
\ome{A}{A}&=&{r\over {s(s-r^2-r^4+s r^4)}} [(r^2-s)
{}~\kappa(\T{A}{B} )d\T{B}{A}
+ r^2 (s-1)~\kappa (\T{C}{B}) d\T{B}{C} \theta^{CA}+\nonumber\\
& &+(-r^2-s r^2 +s +s r^4)~\kappa(\T{C}{B}) d\T{B}{C}
\theta^{AC}],~~~
{}~~~\mbox{no sum on A}~~~
\label{omAA}\\
\ome{A}{B}&=&-s^{-1} {r\over q_{BA}} \kappa (\T{B}{C} )d\T{C}{A}
,~~~~~~~~~~~~~~~~~~~~~~~~~~~~~~~A \not= B
\label{omAB}
\ena

\noi When $s=1$, the classical limit $\ome{A}{B}
\qronelim -\kappa (\T{A}{C})
d\T{C}{B}$  reproduces the familiar formula
$\om=-g^{-1}dg$ for the left-invariant
one-forms on the group manifold.  More generally, for
$s=r^{\alpha}, \alpha \in {\bf C}$ we have :
\eq
\ome{A}{A} \ronelim [ {{2-\alpha} \over {2 (\alpha -1)}}
\sum_B \kappa(\T{A}{B} )d\T{B}{A}
+{\alpha \over {2 (\alpha -1)}} \sum_B \sum_{C \not= A}
\kappa (\T{C}{B})
d\T{B}{C}],~~~\mbox{no sum on A} ,
\en
\noi which shows that the inversion formula (\ref{omAA})
diverges in the classical limit for
$s=r$.
\sk
Due to the bi-invariance of $\tau$ the derivative operator $d$ is
compatible with the actions $\DL$ and $\DR$:
\eq
\DL (da)=(id\otimes d)\D(a)\label{propdaL}
\en
\eq
\DR (da)=(d\otimes id)\D(a),\label{propdaR}
\en
These two properties express the fact that $d$
commutes with the left
and right action of the quantum group, as in the classical case.
\sk

We conclude that the exterior derivative (\ref{defd1})
together with the properties (\ref{Leibniz}),
(\ref{propdaL}) , (\ref{propdaR}) and (\ref{propd})
realize the axioms of a first-order bicovariant differential
calculus
\cite{Wor}.
\sk
\noi{\bf Tensor product}
\sk
The tensor product between elements $\rho,\rhop \in \Ga$
is defined to
have the properties $\rho a\otimes \rhop=\rho
\otimes a \rhop$, $a(\rho
\otimes \rhop)=(a\rho) \otimes \rhop$ and
$(\rho \otimes \rhop)a=\rho
\otimes (\rhop a)$. Left and right actions on
$\Ga \otimes \Ga$ are
defined by:
\eq
\DL (\rho \otimes \rhop)\equiv \rho_1 \rhop_1
\otimes \rho_2 \otimes
\rhop_2,~~~\DL: \Ga \otimes \Ga \rightarrow A
\otimes\Ga\otimes\Ga
\label{DLGaGa}
\en
\eq
\DR (\rho \otimes \rhop)\equiv \rho_1 \otimes \rhop_1
 \otimes \rho_2
\rhop_2,~~~\DR: \Ga \otimes \Ga \rightarrow
\Ga\otimes\Ga\otimes A
\label{DRGaGa}
\en
\noi where  $\rho_1$, $\rho_2$, etc., are defined by
\eq
\DL (\rho) = \rho_1 \otimes \rho_2,~~~\rho_1\in A,~\rho_2\in \Ga
\en
\eq
\DR (\rho) = \rho_1 \otimes \rho_2,~~~\rho_1\in \Ga,~\rho_2\in A.
\en
\noi The extension to $\Ga^{\otimes n}$ is straightforward.
\sk

\noi {\bf Exterior product}
\sk

The  exterior product of one-forms is consistently defined as:
\eq
\ome{A_1}{A_2} \we \ome{D_1}{D_2}
\equiv \ome{A_1}{A_2} \otimes \ome{D_1}{D_2}
- \RRhat{A_1}{A_2}{D_1}{D_2}{C_1}{C_2}{B_1}{B_2}
\ome{C_1}{C_2} \otimes \ome{B_1}{B_2} \label{exteriorproduct}
\en
\noi where the $\Lambda$ tensor is given by:
\eq
\LL{A_1}{A_2}{D_1}{D_2}{C_1}{C_2}{B_1}{B_2}
\equiv \ff{A_1}{A_2B_1}{B_2} (\MM{C_1}{C_2D_1}{D_2})
=
\en
\eq
d^{F_2} d^{-1}_{C_2} \R{F_2B_1}{C_2G_1} \Rinv{C_1G_1}{E_1A_1}
    \Rinv{A_2E_1}{G_2D_1} \R{G_2D_2}{B_2F_2} \label{Lambda}
\en
This matrix satisfies the characteristic equation:
\eq
(\La+r^2 I)~(\La+r^{-2}I)~(\La-I)=0 \label{Laeigen}
\en
\noi due to the Hecke condition (\ref{Hecke}).
For simplicity
we will at times use the adjoint indices $i,j,k,...$
with ${}^i={}_A^{~B},~{}_i={}^A_{~B}$. Then
(\ref{Laeigen}) applied to $\om^r \otimes \om^s$
yields:
\eqa
\lefteqn{(\L{ij}{kl} + r^2 \de^{i}_{k} \de^{j}_{l})
(\L{kl}{mn} + r^{-2} \de^{k}_{m} \de^{l}_{n})
(\L{mn}{rs} -  \de^{m}_{r} \de^{n}_{s})\om^r \otimes \om^s=}
\nonumber\\
& & (\L{ij}{kl} + r^2 \de^{i}_{k} \de^{j}_{l})
(\L{kl}{mn} + r^{-2} \de^{k}_{m} \de^{l}_{n})
\om^m \we \om^n = 0
\ena
\noi and it is easy to see that the last equality
 can be rewritten as
\eq
\om^i \we \om^j = - \Z{ij}{kl} \om^k \we \om^l \label{commom}
\en
\eq
\Z{ij}{kl} \equiv {1\over {r^2 + r^{-2}}} [\L{ij}{kl} + \Linv
{ij}{kl}]. \label{defZ}
\en
\noi {\sl Note: } The inverse of $\Lambda$ always
exists, and is given by
\eqa
\lefteqn{\RRhatinv{A_1}{A_2}{D_1}{D_2}{B_1}{B_2}{C_1}{C_2}=
\ff{D_1}{D_2B_1}{B_2}
(\T{A_2}{C_2} \km (\T{C_1}{A_1}))=}\nonumber\\
& & \R{F_1B_1}{A_1G_1} \Rinv{A_2G_1}{E_2D_1} \Rinv{D_2E_2}{G_2
C_2} \R{G_2C_1}{B_2F_1} (d^{-1})^{C_1} d_{F_1}
\label{Lambdainv}
\ena
\sk
\noi {\bf Exterior differential on $\Gamma^{\we n} $}
\sk

Having the exterior product we can define the exterior
differential on $\Ga$:
\eq
d~:~\Gamma \rightarrow \Gamma \we \Gamma
\en
\eq
d (a_k db_k) = da_k \we db_k \label{defdonga}
\en
\noi which can easily be extended to
$\Gamma^{\we n}$ ($d: \Gamma^{\we n}
\rightarrow \Gamma^{\we (n+1)}$, $\Gamma^{\we n}$ being
defined as in the
classical case but with the quantum permutation
operator $\La$ \cite{Wor}). The definition (\ref{defdonga})
is equivalent to the following :
\eq
d\theta \equiv \rinv [\tau \we \theta - (-1)^k \theta \we \tau],
\label{defdgen}
\en
\noi where $\theta \in \Ga^{\we k}$,
and has the following properties:
\eq
d(\theta \we \thetap)=d\theta \we \thetap +
(-1)^k \theta \we d\thetap
\label{propd1}
\en
\eq
d(d\theta)=0\label{propd2}
\en
\eq
\DL (d\theta)=(id\otimes d)\DL(\theta)\label{propd3}
\en
\eq
\DR (d\theta)=(d\otimes id)\DR(\theta),\label{propd4}
\en
\noi where $\theta \in \Ga^{\we k}$, $\thetap \in \Ga^{\we n}$.
\sk
\noi {\bf Cartan-Maurer equations}
\sk
The $q$-Cartan-Maurer equations are found by using
(\ref{defdgen}) in computing $d\ome{C_1}{C_2}$:
\eq
d\ome{C_1}{C_2}= \rinv (\ome{B}{B} \we \ome{C_1}{C_2} +
 \ome{C_1}{C_2} \we
\ome{B}{B}) \equiv
-\onehalf \cc{A_1}{A_2}{B_1}{B_2}{C_1}{C_2}
{}~\ome{A_1}{A_2} \we \ome{B_1}{B_2} \label{CartanMaurer}
\en
\noi with:
\eq
\cc{A_1}{A_2}{B_1}{B_2}{C_1}{C_2}=-{2\over{r-r^{-1}}}
 [\de^{A_1}_{C_1}
\de^{C_2}_{A_2} \de^{B_1}_{B_2}-{1\over {r^2+r^{-2}}}
(\de^{A_1}_{C_1} \de^{C_2}_{A_2} \de^{B_1}_{B_2}+\LL{B}{B}
{C_1}{C_2}{A_1}{A_2}{B_1}{B_2})] \label{cc}
\en
To derive this formula we have used the flip operator $Z$
on $\ome{B}{B} \we \ome{C_1}{C_2}$.
\sk
\noi {\bf q-Lie algebra}
\sk
Finally, we recall that the $\chi$ operators close
on the q-Lie algebra:
\eq
\chi_i \chi_j - \L{kl}{ij} \chi_k \chi_l = \C{ij}{k} \chi_k
\label{bico1}
\en
\noi where the $q$-structure constants are given by $\C{jk}{i}=
\chi_k(\M{j}{i})$ or explicitly:
\eq
\CC{A_1}{A_2}{B_1}{B_2}{C_1}{C_2} =\rinv [- \de^{B_1}_{B_2}
\de^{A_1}_{C_1} \de^{C_2}_{A_2} +
\LL{B}{B}{C_1}{C_2}{A_1}{A_2}{B_1}{B_2}]. \label{CC}
\en
Comparing with (\ref{cc}) we find the relation between
the structure constants $C$ appearing in the Cartan-Maurer
equations and the structure constants ${\bf C}$ of the $q$-Lie
algebra of the tangent vectors in eq. (\ref{bico1}):
\eq
\cc{A_1}{A_2}{B_1}{B_2}{C_1}{C_2}={2\over {r^2+r^{-2}}}
[-(r-r^{-1}) \de^{B_1}_{B_2} \de^{A_1}_{C_1} \de^{C_2}_{A_2}
+\CC{A_1}{A_2}{B_1}{B_2}{C_1}{C_2}]
\en
For $\qrone$ the structure constants $C$ and $\bf C$
coincide.
\sk
More in general the $\chi$ and $f$ operators close on the algebra
(\ref{bico1}) and
\eq
\L{nm}{ij} \f{i}{p} \f{j}{q} = \f{n}{i} \f{m}{j} \L{ij}{pq}
\label{bico2}
\en
\eq
\C{mn}{i} \f{m}{j} \f{n}{k} + \f{i}{j} \chi_k= \L{pq}{jk} \chi_p
\f{i}{q} + \C{jk}{l} \f{i}{l} \label{bico3}
\en
\eq
\chi_k  \f{n}{l}=\L{ij}{kl} \f{n}{i} \chi_j,  \label{bico4}
\en
This algebra is {\sl sufficient} to define
a bicovariant differential calculus on $A$
(see e.g. \cite{Bernard}), and will be called
``bicovariant algebra" in the sequel.
By applying the relations defining
the bicovariant algebra (called also
``bicovariance conditions")  to the element $\M{r}{s}$
we can express them
in the adjoint representation:
\eqa
& & \C{ri}{n} \C{nj}{s}-\L{kl}{ij} \C{rk}{n} \C{nl}{s} =
\C{ij}{k} \C{rk}{s}
{}~~\mbox{({\sl q}-Jacobi identities)} \label{bicov1}\\
& & \L{nm}{ij} \L{ik}{rp} \L{js}{kq}=\L{nk}{ri} \L{ms}{kj}
\L{ij}{pq}~~~~~~~~~\mbox{(Yang--Baxter)} \label{bicov2}\\
& & \C{mn}{i} \L{ml}{rj} \L{ns}{lk} + \L{il}{rj} \C{lk}{s} =
\L{pq}{jk} \L{is}{lq} \C{rp}{l} + \C{jk}{m} \L{is}{rm}
\label{bicov3}\\
& & \C{rk}{m} \L{ns}{ml} = \L{ij}{kl} \L{nm}{ri} \C{mj}{s}
\label{bicov4}
\ena
\sk
Using the definitions (\ref{defchi}) and (\ref{defff}) it is
not difficult to find the co-structures
on the functionals $\chi$ and $f$:
\eqa
& & \Dp (\chi_i)=\chi_j
     \otimes \f{j}{i} + \epsi \otimes \chi_i \label{copchi}\\
& & \ep(\chi_i)=0 \label{couchi}\\
& & \kp(\chi_i)=-\chi_j \kp(\f{j}{i}), \label{coichi}
\ena
\eqa
& & \Dp (\f{i}{j})=\f{i}{k} \otimes \f{k}{j}   \label{copf}\\
& & \ep (\f{i}{j}) = \del^i_j  \label{couf}\\
& & \kp (\f{k}{j}) \f{j}{i}= \de^k_i \epsi =
\f{k}{j} \kp (\f{j}{i})
\label{coif}
\ena
\noi Note
that in the $r,q,s\rightarrow 1$ limit $\f{i}{j}
\rightarrow \de^i_j \epsi$, i.e. $\f{i}{j}$ becomes
proportional to the
identity functional and formula (\ref{omea}),
becomes trivial, e.g. $\om^i a = a\om^i$ [use $\epsi * a=a$].
\sk

%%%%%%%%%%%%%%%%%%%%%%%%%%%%%%

\sect{The quantum group $IGL_{q,r} (N)$}

%%%%%%%%%%%%%%%%%%%%%%%%%%%%%%

The $q$-inhomogeneous group $\IGLqrN$  is freely generated by the
non-commuting matrix elements $T^A{}_B~ [A=(0,a) ; a: 1,..N]$, the
identity $I$ and the inverse $\xi$ of the $q$-determinant of $T$ as
defined in (\ref{qrdet}), modulo the relations:
\eq
T^0{}_a=0 \label{proietto}
\en
and the relations:

\eqa
& &\R{ab}{ef} \T{e}{c} \T{f}{d}=\T{b}{f} \T{a}{e} \R{ef}{cd}
\label{RTTIGL1}\\
& &\R{ab}{ef} \T{e}{c} x^f= {q_{0c} \over r}
 x^b \T{a}{c} \label{RTTIGL2} \\
& &\R{ab}{ef} x^e x^f=r x^b x^a \label{RTTIGL3}\\
& &q_{0a}\T{a}{c} u=q_{0c} u\T{a}{c} \label{RTTIGL4}\\
& &q_{0a} x^a u = u x^a \label{RTTIGL5}
\ena
\noi where $x^a \equiv \T{a}{0}$ and $u\equiv T^0{}_0$.

It is not difficult to check that this algebra,
endowed with the coproduct $\D$, the counit
$\epsi$ and the coinverse
$\kappa$ defined by :
\eqa
\D(\T{A}{B})=\T{A}{C} \otimes \T{C}{A} ;
{}~ & \epsi(\T{A}{B})=\de^A_B
;~ & \kappa(T)=T^{-1} \label{TperT}\\
\D(\xi)=\xi\otimes \xi ;~ & \epsi(\xi)=1 ;
{}~ & \kappa(\xi)=\det T
\label{xiperxi} \\
\D(I)=I\otimes I ;~ & \epsi(I)=1 ;~ & \kappa(I)=I
\label{IperI}
\ena

\noi where the quantum inverse of $T^A{}_B$ is given
by $\Ti{A}{B}=\xi~\pp{AB}{0,N}~ t_B^{~A} $
 [see eq. (\ref{defPi}):  $t_B{}^A$ is the
quantum minor ], is a Hopf algebra. The proof goes
as in uniparametric case (see the second ref. of
\cite{inhom}).
\sk
In the commutative limit it is the algebra of
functions on $IGL(N)$
plus the dilatation $T^0{}_0$.
\sk
Relations (\ref{TperT})-(\ref{IperI}) explicity read:
\eqa
& &\D(\T{a}{b})=\T{a}{c}
\otimes \T{c}{b},~~\D (I)=I\otimes I,\label{DT}\\
& & \D(x^a)=\T{a}{b} \otimes x^b + x^a \otimes u \label{Dx}\\
& & \D(u)=u\otimes u,~~\D(\xi)=\xi\otimes \xi \label{Du}\\
& & \D(\detqrTab)=\detqrTab \otimes \detqrTab
\label{Ddet}
\ena
\eqa
& & \epsi(\T{a}{b})=\de^a_b,~~\epsi (I)=1,\label{epsiT}\\
& & \epsi(x^a)=0 \label{epsix}\\
& & \epsi(u)=\epsi(\xi)=1 \label{epsiu}\\
& & \epsi(\detqrTab)=1 \label{epsidet}
\ena
\eqa
& & \kappa(\T{a}{b})=\Ti{a}{b}=\xi u ~\pp{ab}{1,N} t_b^{~a}\\
& & \kappa(I)=I, \label{kappaT}\\
& & \kappa(x^a)=-\kappa(\T{a}{b}) x^b \kappa (u) \label{kappax}\\
& & \kappa(u)= \detqrTab ~\xi \label{kappau}\\
& & \kappa(\xi)=u ~\detqrTab,~~\kappa(\detqrTab)=\xi u
\label{kappadet}
\ena
\noi where for completeness we have included the
expressions for the
$q$-determinant of $T$. Note that
$\kappa (u) u = I = u \kappa(u)$.

\sk
This procedure is very similar to that
discussed for $\GLqrNo$
in Section 2: indeed both these Hopf
algebrae are obtained from
the algebra freely generated by
$T^A{}_B, I,\Xi\mbox{ or } \xi$ through the
introduction of moduli relations i.e.
as quotients of suitable
two-sided ideals: the one generated by the
$RTT$ relations in the
$\GLqrNo$ case, and the one generated by the
(\ref{proietto})-(\ref{RTTIGL5}) relations
in the $\IGLqrN$ case.
\sk
We now rederive the quantum group $\IGLqrN$ as a
quotient of $\GLqrNo$: all Hopf algebra properties
of $\IGLqrN$ will descend from
those of $\GLqrNo$. The formalism
employed will be useful in the next
Section to deduce the differential
calculus on $\IGLqrN$ from the one on $\GLqrNo$.
\sk
We start from the observation that the $R$-matrix of $\GLqrNo$ can
be written as  ({\small A=(0,a)}):
\eq
\R{AB}{CD}=\left( \begin{array}{cccc}
                  r&0&0&0\\0&{r\over q_{0b}}\de^b_d&0&0\\
                 0&(r-r^{-1})\de^a_d &{q_{0a}\over r} \de^a_c&0\\
                  0&0&0&\R{ab}{cd}\\
                \end{array} \right) \label{RGLqrNo}
\en
\noi where $\R{ab}{cd}$ is the $R$-matrix of $GL_{q,r}(N)$, and
the indices {\small AB} are
ordered as {\small $00,0b,a0,ab$}.

It is apparent that the $\GLqrNo$ $R$ matrix
 contains the information on $\GLqrN$.
We will show that it also contains the information
about the quantum group $\IGLqrN$.

In the index notation $A=(0,a)$ the $RTT$ relations explicity
read :
\eqa
& &\R{ab}{ef} \T{e}{c} \T{f}{d}=\T{b}{f} \T{a}{e} \R{ef}{cd}
\label{exp0}\\
& &T^{a}{}_{c}T^{b}{}_{0} = \frac{q_{ab}}
{q_{c0}}T^{b}{}_{0}T^{a}{}_{c}
\label{exp1} \\
& &T^{a}{}_{0}T^{b}{}_{d} =
\frac{q_{ab}}{q_{0d}}T^{b}{}_{d}T^{a}{}_{0}
+\frac{r}{q_{0d}}(r-r^{-1})T^{a}{}_{d}T^{b}{}_{0}
\label{exp2} \\
& &T^{a}{}_{c}T^{0}{}_{d} = \frac{q_{a0}}
{q_{cd}}T^{0}{}_{d}T^{a}{}_{c}
\label{exp3} \\
& &T^{0}{}_{c}T^{b}{}_{d} =
\frac{q_{0b}}{q_{cd}}T^{b}{}_{d}T^{0}{}_{c}
+\frac{r}{q_{cd}}(r-r^{-1})T^{0}{}_{d}T^{b}{}_{c}
\label{exp4} \\
& &\T{a}{0} \T{b}{0}=q_{ab} \T{b}{0} \T{a}{0} \\
& &T^{0}{}_{c}T^{b}{}_{0} = \frac{q_{0b}}{q_{c0}}T^{b}
{}_{0}T^{0}{}_{c}
\label{exp5} \\
& &\T{0}{c} \T{0}{d}=q_{dc} \T{0}{d} \T{0}{c}\\
& &T^{0}{}_{0}T^{b}{}_{d} =
\frac{q_{0b}}{q_{0d}}T^{b}{}_{d}T^{0}{}_{0}
+\frac{r}{q_{0d}}(r-r^{-1})T^{0}{}_{d}T^{b}{}_{0}
\label{exp6} \\
& &T^{0}{}_{0}T^{b}{}_{0} = {q_{0b}}T^{b}{}_{0}T^{0}{}_{0}
\label{exp7} \\
& &T^{0}{}_{0}T^{0}{}_{d} = {q_{d0}}T^{0}{}_{d}T^{0}{}_{0}
\label{exp8}
\ena
where $a<b$ and $c<d$.
\sk
Consider now in $\GLqrN$ the space $H$ of all sums of
monomials containing at least an element
of the kind $T^0{}_a$
(i.e. $H$ is the ideal in $\GLqrNo$ generated
by the elements $T^0{}_a$
as we will see).
Notice that
$T^{0}{}_{0}T^{b}{}_{d} -\frac{q_{0b}}
{q_{0d}}T^{b}{}_{d}T^{0}{}_{0}$
is an element of $H$ because of relation (\ref{exp6}).
\sk
\noi{\sl Proposition 1.} ${}~~~~H$ is the space of
all sums of monomials of
the kind $T^0{}_ba^b$ with $a^b \in \GLqrNo$.

\noi {\sl Proof:} every monomial in a generic
element  $h$ of  $H$ contains at least a
factor $T^0{}_a$. Use the $RTT$ relations (\ref{exp3}),
(\ref{exp4}), (\ref{exp5}) and (\ref{exp8}) to see that it can be
shifted step by step to the left until we have
a sum of monomials $h=T^0{}_b a^b.$
\sk
A similar proof holds for the
\sk
\noi{\sl Proposition 2.} ${}~~~~H$ is the space of all
sums of monomials of
the kind $a^bT^0{}_b$ with $a^b \in
\GLqrNo$.
\sk
We now prove that $H$ is a Hopf ideal, i.e. an ideal
in the $\GLqrNo$
algebra that is also compatible with the
co-structures of $\GLqrNo$;
this allows to structure ${\GLqrNo}/{H}$ as a Hopf algebra.
\sk

\noi {\sl Theorem 1:} The space $H$ is a Hopf ideal
in $\GLqrNo$ that is:
\sk
\noi ${}~~{}$i) ${}~~{}$ $H$ is a two-sided ideal in $\GLqrNo$
\sk
\noi ${}~{}$ii) ${}~~{}$ $H$ is a co-ideal i.e.
\eq
{}~~~~~\DN(H)\subseteq H \otimes \GLqrNo +
\GLqrNo\otimes H ~;~~ \eN (H) =0
\label{coideal}
\en
\noi ${}{}$iii) ${}~~{}$ $H$ is compatible with
$~\kN$ :
%~\GLqrNo\rightarrow\GLqrNo$
\eq
\kN(H)\subseteq H~.\label{Hideal}
\en

\noi {\sl Proof:}
\sk
\noi ${}~~{}$i) $H$ is trivially a subalgebra of $\GLqrNo$.
It is a right and left ideal since $\forall h\in H,\/ \forall a\in
\GLqrNo ~ha\in H \mbox{ and }
 ah\in H.$
This follows immediately from the definition of $H$ as sums of
monomials containing at least a factor $T^0{}_a$.
$H$ is the
ideal in $\GLqrNo$ generated by the elements $T^0{}_a$.
\sk
\noi ${}~{}$ii)
By virtue of {\sl Proposition 1}, $\forall h \in H$:
\eqa
\DN(h) & = & \DN(T^0{}_ba) = (T^0{}_c  \otimes
T^c{}_b + T^0{}_0 \otimes
T^0{}_b)[a{}_{1}\otimes a{}_{2}] \\
       & = & \underbrace{T^0{}_c a{}_{1}}_{\in \textstyle H}
\otimes T^c{}_b
a{}_{2} + T^0{}_0 a{}_{1} \otimes  \underbrace{T^0{}_b
a{}_{2}}_{\in \textstyle H}
\ena
\noi with the symbolic notation $\D (a) \equiv a_1 \otimes a_2$.
Moreover
\eq
\eN(h) = \eN(T^0{}_b a) = \eN(T^0{}_b)\eN(a) = 0~.
\en
\noi ${}{}$iii)
\eq
\kN(T^0{}_b) =\Xi~\pp{ob}{0,N}~ t_b^{~0}
\en
where $\pp{ob}{0,N}$ is defined in (\ref{defPi})
and  it is easy to see that the quantum minor
$t_b{}^0 \in H$ since it is
the determinant of a matrix that has elements
$T^0{}_a$ in the first row.
By {\sl Proposition 1} and {\sl Proposition 2} we have
\eq
\kN(h)=\kN(T^0{}_ba)=\kN(a)\kN(T^0{}_b) \in H~,
\en
\noi Q.E.D.
\sk
We now consider the quotient
\eq
\frac{\GLqrNo}{H}~,
\label{quotient}
\en
\noi i.e. the space of all equivalence
classes $\{a\} ,~a\in \GLqrNo $
where $\{a\}$
contains all those elements of $\GLqrNo$
that differ from $a$ by an
element of $H$; in particular $\{T^0{}_a\}=\{0\}$.  Let us
introduce the canonical
projection
\eqa
P ~:~~ & \GLqrNo\longrightarrow & \GLqrNo/{H}  \\
       & ~~~~~~~~~~~a\longmapsto & P(a)\equiv \{a\}~~
\ena
Any element of ${\GLqrNo/H}$ is of the form $P(a)$. Also, $P(H)=0$,
i.e. $H=Ker(P)$.
\sk
Since $H$ is a two-sided ideal, ${\GLqrNo/H}$ is an algebra with
the following sum and products:
\eq
P(a)+P(b)\equiv P(a+b) ~;~~ P(a)P(b)\equiv P(ab) ~;~~
\mu P(a)\equiv P(\mu a),~~~\mu \in \mbox{\bf C} \label{iglalgebra}
\en
We will use the following notation:
\eq
x^a\equiv P(T^a{}_0) ~;~~ u\equiv P(T^0{}_0) ~;~~ \xi\equiv P(\Xi)
\en
and with abuse of symbols:
\eq
T^a{}_b\equiv P(T^a{}_b) ~;~~ I\equiv P(I) ~;~~ 0\equiv P(0)
\en
notice that $P(T^0{}_a)=P(0)=0$.
Using (\ref{iglalgebra}) it is easy to show that
$T^a{}_b , ~x^a , ~u, ~\xi$ and $I$
generate the algebra ${\GLqrNo/H}$. Moreover from
the $RTT$  relations
$R_{12}T_1T_2=T_2T_1R_{12}$ in $\GLqrNo$  we find the ``$P(RTT)$"
relations in ${\GLqrNo/H}$:
\eq
P(R_{12}T_1T_2)=P(T_2T_1R_{12}) ~~~ i.e. ~~~
R_{12}P(T_1)P(T_2)=P(T_2)P(T_1)R_{12}  \label{PRTT}
\en
that are explicity given in  (\ref{RTTIGL1})-(\ref{RTTIGL5}).
\sk
Since H is a Hopf ideal then ${\GLqrNo/H}$ is also a Hopf
algebra with co-structures:
\eq
\D(P(a))\equiv (P\otimes P)\DN(a) ~;~~ \epsi(P(a))
\equiv\eN(a) ~;~~
\kappa(P(a))\equiv P(\kN(a)) \label{co-igl}
\en
Indeed (\ref{coideal}) and (\ref{Hideal}) ensure
that $\D ,~\epsi ,$
and $\kappa$ are well defined. For example
\eq
(P\otimes P)\DN(a) = (P\otimes P)\DN(b)~~
\mbox{ if }~~ P(a)=P(b) ~.
\en
In order to prove the Hopf algebra axioms of Appendix A for
$\D,~\epsi,~\kappa$ we just have to project
those for $\DN,~\eN,~\kN~.$
For example, the first axiom is proved by applying
$P\otimes P \otimes P$ to $(\DN \otimes id)\DN(a) =
 (id \otimes \DN)\DN(a)$. The other axioms are proved
in a similar way.
\sk
Notice that on the generators $T^a{}_b , ~x^a , ~u,
{}~\xi$ and $I$ the
co-structures (\ref{co-igl}) act as in
(\ref{TperT})-(\ref{IperI}).
\sk
In conclusion: the elements $T^a{}_b , ~x^a , ~u,
{}~\xi$ and $I$
generate the Hopf algebra ${\GLqrNo/H}$ and satisfy the
``$P(RTT)$'' commutation rules
(\ref{RTTIGL1})-(\ref{RTTIGL5}). The
co-structures act on them exactly as the co-structures defined in
(\ref{TperT})-(\ref{IperI}). Therefore the quotient
${\GLqrNo/H}$ is the $q$-inhomogeneous group defined at the
beginning of this section:
\eq
\IGLqrN = \frac{\GLqrNo}{H}  ~.
\en
The canonical projection $P ~:~~ \GLqrNo \rightarrow
\IGLqrN$ is an
epimorphism between these two Hopf algebrae.
\sk
\noi{\sl Note 1:} From the commutations
(\ref{RTTIGL4}) - (\ref{RTTIGL5})
 we see that
one can set $u=I$ only when $q_{0a}=1$ for all $a$.
\sk
\noi{\sl Note 2:}  $P(\detqrTAB)=u~\detqrTab$
is central in $\IGLqrN$
only when $Q_A=1$, {\small {\it A=0,1,..N}}
(apply the projection $P$ to eq. (\ref{detcomm})). Note
that here we have $Q_A \equiv
\prod_{C=0}^N ({q_{CA}\over r})$.
\sk
\noi{\sl Theorem 2:} The centrality of $u$ is incompatible with
the centrality of $\detqrTab$.
\sk
\noi{\sl Proof:}  Suppose that
$q_{0a}=1$ so that $u$ is central.
Then the centrality of $\detqrTab$
is equivalent to the centrality
of $P(\detqrTAB)$ and requires $Q_A=1$ (previous Note);
in particular $Q_0 \equiv
\prod_{c=1}^N {r\over q_{0c}}=1$, which cannot be since
for $q_{0a}=1$ we find $Q_0=r^N$.  Q.E.D.
\sk
We end this Section by giving the commutations
of $\det \T{a}{b}$ and $\xi$ with all the generators:
\eq
(\det \T{c}{d}) \T{a}{b}={\Qt_a \over \Qt_b} \T{a}{b}
 (\det \T{c}{d}),~~~~~\zeta \T{a}{b}={\Qt_b \over \Qt_a} \T{a}{b}
 \zeta
\en
\eq
(\det \T{c}{d}) x^a ={\Qt_a \over Q_0} x^a (\det \T{c}{d}),
{}~~~~~\zeta x^a ={Q_0 \over \Qt_a} x^a \zeta
\en
\eq
(\det \T{c}{d}) u = u (\det \T{c}{d}),
{}~~~~~\zeta u =u \zeta
\en
\noi where here $Q_a \equiv
\prod_{c=1}^N ({q_{ca}\over r})$ and $\zeta$ is the
inverse of $\det \T{c}{d}$, i.e. $\zeta \equiv u \xi$.
We see that the commutations of $\det \T{c}{d}$ with
$\T{a}{b}$ are the correct ones for $\GLqrN$ (i.e. are
identical to the ones deduced in Section 2).
In the standard uniparametric case $Q_a=1$, and
the $q$-determinant $\det \T{c}{d}$ becomes central
(and likewise $\zeta$),
provided that also $Q_0=1$.

%%%%%%%%%%%%%%%%%%%%%%%%%%%%%%%%%%%%

\sect{The differential calculus of $\IGLqrN$}

%%%%%%%%%%%%%%%%%%%%%%%%%%%%%%%%%%%%

In this Section we present  a
bicovariant differential calculus on $\IGLqrN$,
based on  the following set of
functionals $f$ and elements $M$ :
\eqa
& &\ff{a_1}{a_2b_1}{b_2}= \kp (\Lp{b_1}{a_1}) \Lm{a_2}{b_2}
\nonumber\\
& &\ff{a_1}{a_2 0}{b_2}= \kp (\Lp{0}{a_1}) \Lm{a_2}{b_2}
\nonumber\\
& &\ff{0}{a_2b_1}{b_2}= \kp (\Lp{b_1}{0}) \Lm{a_2}{b_2} \equiv 0
\nonumber\\
& &\ff{0}{a_2 0}{b_2}= \kp (\Lp{0}{0}) \Lm{a_2}{b_2} \label{fin}
\ena
\eqa
& &\MM{b_1}{b_2a_1}{a_2} = \T{b_1}{a_1} \kappa (\T{a_2}{b_2})
\nonumber\\
& &\MM{b_1}{b_2 0}{a_2} = \T{b_1}{0} \kappa (\T{a_2}{b_2})
\nonumber\\
& &\MM{0}{b_2a_1}{a_2} = 0
\nonumber\\
& &\MM{0}{b_2 0}{a_2} = \T{0}{0} \kappa (\T{a_2}{b_2}) \label{Min}
\ena
The $f$ in (\ref{fin})
 are a subset of the $f$ functionals of
$\GLqrNo$, obtained by restricting  the indices
of  $\f{i}{j}$
%and $\chi_i$
to $i=ab$ and $i=0b$. The third $f$ is identically zero because
of upper triangularity of $L^+$, i.e. $\Lp{b_1}{0}=0$.

The elements $M\in \IGLqrN$ in (\ref{Min})
 are obtained with the same
restriction on the adjoint indices, and by projecting
with $P$. The effect of the projection is to replace
the coinverse in $\GLqrNo$, i.e. $\kappa_{N+1}$ ,
with the coinverse $\kappa$ of $\IGLqrN$ (see the
last of (\ref{co-igl})). The
third element in (\ref{Min}) becomes zero because of $P$.
\sk
{\sl Theorem 1:} the functionals in (\ref{fin})
vanish when applied to elements of $Ker(P) \subset
\IGLqrN$.
\sk
{\sl Proof:} first one checks directly that the
functionals (\ref{fin}) vanish
when applied to $\T{0}{b}$.  This extends to any element
of the form $\T{0}{b} a$ ($a \in A$), i.e. to any element of
$Ker(P)$, because of the property (\ref{propf1}) and
the vanishing of the functionals in
(\ref{fzero}). Q.E.D.
\sk
The theorem states that the $f$ functionals are
well defined on the quotient $\IGLqrN = \GLqrNo / Ker(P)$,
in the sense that the ``projected" functionals
\eq
\fb : \IGLqrN \rightarrow \Cb,~~
\fb (P(a)) \equiv f(a)~ , ~~~\forall a \in \GLqrNo \label{deffb}
\en
are well defined.
Indeed if $P(a)=P(b)$, then $f(a)=f(b)$ because
$f(Ker(P))=0$. This holds for any
 functional $f$ vanishing on $Ker(P)$, not only for
the $\f{i}{j}$ functionals.
\sk
The product $fg$ of  two generic functionals
vanishing on $KerP$ also vanishes
on $KerP$, because $KerP$ is a co-ideal (see
Theorem 1 in Section 5): $fg(KerP)=(f\otimes g)\DN (KerP)=0$.
Therefore $\overline{fg}$ is well defined on $\IGLqrN$, and
\eq
\overline{fg}[P(a)]=fg(a)=(f\otimes g)\D_{N+1}(a)=
(\fb P \otimes \gb P) \DN (a)=
(\bar{f}\otimes \bar{g}) \D(P(a)) \equiv
 \bar{f} \bar{g}[P(a)] \label{fgbar}
\en
\noi (use the first of (\ref{co-igl}))
so that the product of $\bar{f}$ and $\bar{g}$ involves
the coproduct $\D$ of $\IGLqrN$.
\sk
There is a natural way to introduce a coproduct on the $\fb$'s :
\eq
\D'\fb[P(a)\otimes P(b)]\equiv\fb[P(a)P(b)]=\fb[P(ab)]
=f(ab)=\D'_{N+1}f(a\otimes b) ~.
\en
It is also easy to show that
\eq
\D'\fb^i{}_j = \fb^i{}_k\otimes \fb^k{}_j ~~\mbox{ i.e. }~
\fb^i{}_j[P(a)P(b)] = \fb^i{}_k[P(a)]\fb^k{}_j[P(b)]
\label{cofb}
\en
with $i,j,k$ running over the restricted set of indices $ab, 0b$.
Indeed due to
\eq
\ff{0}{A_2 b_1}{B_2} \equiv 0,~~\ff{A_1}{0 B_1}{b_2} \equiv 0
\label{fzero}
\en
\noi (a consequence of upper and lower triangularity of $L^+$ and
$L^-$ respectively ), formulae (\ref{copf}) and (\ref{propf1})
involve only the $f^i{}_j$ listed in (\ref{fin}), which
 annihilate $KerP$. Then
\eq
\fb^i{}_j[P(a)P(b)]=\fb^i{}_j[P(ab)]=f^i{}_j(ab)
=f^i{}_k(a)f^k{}_j(b)=\fb^i{}_k[P(a)]\fb^k{}_j[P(b)]
\en
and (\ref{cofb}) is proved.

With abuse of notations we
will simply write $f$ instead of $\fb$. Then the
$f$  in (\ref{fin}) will be seen as functionals on
$\IGLqrN$.
\sk
{\sl Theorem 2:} the right $A$-module ($A=\IGLqrN$)
 $\Ga$ freely generated
by $\om^i \equiv \ome{a_1}{a_2}, \ome{0}{a_2}$ is
a bicovariant bimodule over $\IGLqrN$
 with right multiplication:
\eq
\om^i a = (\f{i}{j} * a) \om^j,~~a \in \IGLqrN \label{omia}
\en
\noi where the $\f{i}{j}$ are given in (\ref{fin}), the $*$-product
is computed with the co-product $\D$ of $\IGLqrN$,  and
the left and right actions
of $\IGLqrN$ on $\Ga$ are given by
\eqa
& &\DL (a_i \om^i) \equiv \D(a_i) I \otimes \om^i
 \label{DLin}\\
& &\DR (a_i \om^i) \equiv \D(a_i) \om^j \otimes \M{j}{i}
\label{DRin}
\ena
\noi where the $\M{j}{i}$ are given in (\ref{Min}).
\sk
{\sl Proof:} we prove the theorem
by showing  that  the
functionals $f$
and the elements $M$ listed in (\ref{fin}) and
(\ref{Min})
satisfy the properties (\ref{propf1})-(\ref{propM})
(cf. the theorem by Woronowicz discussed in the Section 4).
 It  is straightforward to verify directly that
 the elements $M$
in (\ref{Min}) do satisfy the properties
(\ref{copM}) and (\ref{couM}).
We have already shown that
the functionals $f$ in (\ref{fin}) satisfy
(\ref{propf1}),
and property (\ref{propf2})
obviously also holds for this subset.

Consider now the last property (\ref{propM}).
We know that it holds for $\GLqrNo$.  Take  the free indices
$j$ and $k$  as $ab$ and $0b$, and apply the projection $P$ on
 both members of the equation.
It is an easy matter
to show that  only the $f$'s in (\ref{fin}) and the
$M$'s in (\ref{Min}) enter the sums: this
is due to the vanishing of some $P(M)$ and to (\ref{fzero}).
We still have to prove that the $*$ product
in (\ref{propM})
can be computed via the coproduct $\D$ in
$\IGLqrN$.
Consider the projection of property
 (\ref{propM}), written symbolically as:
\eq
P [ M (f \otimes id) \D_{N+1} (a)]=P [(id\otimes f)
\D_{N+1} (a) M ]~. \label{PMfa}
\en
Now apply the definition (\ref{deffb}) and the first of
(\ref{co-igl}) to rewrite
(\ref{PMfa}) as
\eq
P(M)(\fb\otimes id)\D(P(a))=(id\otimes\fb)\D(P(a))P(M)~.
\en
This projected equation then  {\sl becomes}
 property (\ref{propM})  for the $\IGLqrN$
functionals
$f$ and adjoint elements $M$, with the
correct coproduct $\D$ of $\IGLqrN$.  Q.E.D.
\sk
Using the general formula (\ref{omia}) we can
deduce the $\om, T$ commutations for $\IGLqrN$:
\eqa
& &\ome{a_1}{a_2} \T{r}{s}=s  \Rinv{tb_1}{ca_1}
\Rinv{a_2c}{b_2s} \T{r}{t}\ome{b_1}{b_2}\\
& &\ome{a_1}{a_2} x^r= s {q_{0a_1} \over q_{0a_2}} x^r
\ome{a_1}{a_2} -(r-r^{-1}) {s r\over q_{0a_2}}
\T{r}{a_1} \ome{0}{a_2}\\
& &\ome{a_1}{a_2} u= s {q_{0a_1} \over q_{0a_2}} u~
\ome{a_1}{a_2} \\
& &\ome{a_1}{a_2} \det \T{a}{b}=s^N r^{-2} {Q_{a_1} \over Q_{a_2}}
 (\det \T{a}{b}) \ome{a_1}{a_2}\\
& &\zeta \ome{a_1}{a_2} = s^N r^{-2} {Q_{a_1} \over Q_{a_2}}
 \ome{a_1}{a_2} \zeta\\
& &\ome{0}{a_2} \T{r}{s}=s {r\over q_{0t}}
\Rinv{a_2t}{b_2s} \T{r}{t}\ome{0}{b_2}\\
& &\ome{0}{a_2} x^r= {s\over q_{0a_2}}
 x^r \ome{0}{a_2} \label{Vxcomm}\\
& &\ome{0}{a_2} u= {s\over q_{0a_2}}
 u \ome{0}{a_2}\\
& &\ome{0}{a_2} \det \T{a}{b}=s^N r^{-2} {Q_0 \over Q_{a_2}}
 (\det \T{a}{b}) \ome{0}{a_2}\\
& &\zeta \ome{0}{a_2} =s^N r^{-2} {Q_0 \over Q_{a_2}}
  \ome{0}{a_2}\zeta
\ena
\noi {\sl Note:} $u$ commutes with all
 $\om$ 's only if $q_{0a}=1$ (cf. Note 1 of
Section 5) and $s=1$. This means that
$u=I$ is consistent with the differential calculus
on $IGL_{q_{0a}=1,r}(N)$ only if the additional condition
 $s=1$ is satisfied.
\sk
The 1-form $\tau \equiv \sum_a \ome{a}{a}$
is bi-invariant, as one can check by using
 (\ref{DLin})-(\ref{DRin}).
Then an exterior derivative on $\IGLqrN$ can be defined
 as in eq. (\ref{defd1}), and satisfies the Leibniz rule.
The alternative expression $da=(\chi_i * a) \om^i$
(cf. (\ref{defd2}))
continues to hold, where
\eqa
& &\cchi{a}{b}=\rinv [\ff{c}{c a}{b}-\de^a_b \epsi] \nonumber \\
& &\cchi{0}{b}=\rinv [\ff{c}{c 0}{b}] \label{chiin}
\ena
are the left-invariant vectors dual to the left-invariant
1-forms $\ome{a}{b}$ and $\ome{0}{b}$.
They are functionals on $\IGLqrN$ and as a consequence of
(\ref{cofb}) we have
\eqa
& &\Dp (\cchi{a}{b})=\cchi{c}{d} \otimes \ff{c}{da}{b}+
\epsi \otimes \cchi{a}{b} \label{copchi1}\\
& &\Dp (\cchi{0}{b})=\cchi{c}{d} \otimes \ff{c}{d0}{b}+
\cchi{0}{d} \otimes \ff{0}{d0}{b} +
\epsi \otimes \cchi{0}{b}\label{copchi2}
\ena
\sk
The exterior
derivative on the generators $\T{a}{b}$ is given by
formula (\ref{dTAB}) with lower case indices. For the other
generators we find:
\eqa
& & dx^a=-s {r\over q_{0s}} \T{a}{s} \ome{0}{s}+{{s-1}
\over {r-\rminus}}
x^a \tau \\
& & d u= {{s-1}\over {r-\rminus}} u \tau\\
& & d \xi={{s^{-N-1} r^2-1} \over {r-\rminus}} ~\xi~ \tau
\ena
\noi Moreover:
\eq
 d ( \det \T{a}{b})={{s^N r^{-2}-1}\over{r-\rminus}}
{}~(\det \T{a}{b}) \tau
\en
\eq
 d \zeta={{s^{-N} r^{2}-1}\over{r-\rminus}}
{}~\zeta \tau
\en
\noi ($\zeta \equiv u\xi$). Again we find that $u=I$
implies $s=1$, and $\det \T{a}{b}=\zeta=I$ requires
$s^N r^{-2}=1$.
\sk
Every element $\rho$ of $\Ga$ can be written as
$\rho=a_k db_k$ for some $a_k,b_k$ belonging to
$\IGLqrN$. In fact one has the
same formula as in (\ref{omAA}) for $\ome{m}{n}$,
where all indices now are lower case. For
$\ome{0}{n}$ we find:
\eq
\ome{0}{n}=-{q_{0n}\over {s r}} [\kappa (\T{n}{a})dx^a+
 \kappa (x^n) d u]
\en
Finally, the two properties (\ref{propdaL}) and (\ref{propdaR})
hold also for $\IGLqrN$, because of the bi-invariance
of $\tau=\ome{a}{a}$.
Thus all the axioms for a bicovariant
first order differential calculus on $\IGLqrN$
are satisfied.
\sk
The exterior product of left-invariant one-forms
is defined as
\eq
\om^i \we \om^j\equiv \om^i \otimes \om^j - \L{ij}{kl}
\om^k \otimes \om^l
\en
\noi where
\eq
\L{ij}{kl}=\f{i}{l} (\M{k}{j})
\en
This $\Lambda$ tensor can in fact
be obtained from
the one of $\GLqrNo$ by restricting its indices to the
subset $ab, 0b$. This is true because
when $i,l=ab$ or $0b$ we have
$\f{i}{l} (Ker P)=0$ so that $\f{i}{l}$ is
well defined on $\IGLqrN$, and we can write
$\f{i}{l} (\M{k}{j})=\fb^i_{~l} [P(\M{k}{j})]$
(see discussion after
Theorem 1).
The non-vanishing components of $\Lambda$ read:
\eqa
& &\LL{a_1}{a_2}{d_1}{d_2}{c_1}{c_2}{b_1}{b_2}=
d^{f_2} d^{-1}_{c_2} \R{f_2b_1}{c_2g_1} \Rinv{c_1g_1}{e_1a_1}
    \Rinv{a_2e_1}{g_2d_1} \R{g_2d_2}{b_2f_2} \label{L1}\\
& &\LL{0}{a_2}{d_1}{d_2}{c_1}{c_2}{0}{b_2}={q_{0c_2}\over
   q_{0c_1}} \Rinv{a_2 c_1}{g_2 d_1} \R{g_2d_2}{b_2c_2}
   \label{L2}\\
& &\LL{a_1}{a_2}{0}{d_2}{c_1}{c_2}{0}{b_2}=-(r-r^{-1})
{q_{0c_2}\over
   q_{0a_2}} \de^{c_1}_{a_1}\R{a_2d_2}{b_2c_2}
   \label{L3}\\
& &\LL{a_1}{a_2}{0}{d_2}{0}{c_2}{b_1}{b_2}={q_{0a_1}\over
   q_{0a_2}} d^{f_2} d^{-1}_{c_2} \R{f_2 b_1}{c_2 a_1}
   \R{a_2d_2}{b_2f_2}
   \label{L4}\\
& &\LL{0}{a_2}{0}{d_2}{0}{c_2}{0}{b_2}={q_{0c_2}\over
   q_{0a_2}} r^{-1} \R{a_2d_2}{b_2c_2}
   \label{L5}
\ena
These components still satisfy the characteristic
equation (\ref{Laeigen}), because the $\Lambda$ tensor
of $\GLqrNo$ does satisfy this equation, and
if the free adjoint indices are taken as
$ab$, $0b$, only the components in
(\ref{L1})-(\ref{L5}) enter in (\ref{Laeigen}).
To prove this, consider $\L{ij}{kl}$
with $k,l$  of the type  $ab$
or $0b$ and observe that  it vanishes
unless also $i,j$ are of the type $ab, 0b$.
(This can be checked
directly via the formula (\ref{Lambda})).
Then equations (\ref{commom}) and (\ref{defZ}) hold also
for the $\om$'s of $\IGLqrN$.

Note that $\La^{-1}$ tensor of $\IGLqrN$ can be obtained
by specializing the indices in the $\La^{-1}$ tensor
of $\GLqrNo$ given in (\ref{Lambdainv}), as we did
for $\La$. The reader can convince himself of this
by  i) observing that the $\Linv{ij}{kl}$ tensor of (\ref{Lambdainv})
also vanishes when $k,l$ = $ab$ or $0b$ and $i,j$ are not
of the type $ab, 0b$; ii) considering the equation
$\Linv{ij}{rs} \L{rs}{kl}=\de^i_k \de^j_l$ for $k,l$ = $ab$ or
$0b$.
\sk
The exterior differential on $\Ga^{\we n}$
can be defined as in Section 4 (eq. (\ref{defdgen})),
and satisfies all the properties (\ref{propd1})-(\ref{propd4}).
As for $\GLqrNo$ the last two hold because of the
bi-invariance of $\tau$.
\sk
The Cartan-Maurer equations are
\eq
d\om^i=\rinv (\tau \we \om^i+\om^i \we \tau)=
-\onehalf\c{jk}{i} \om^j \we \om^k
\en
\noi with
\eqa
& &\cc{a_1}{a_2}{b_1}{b_2}{c_1}{c_2}={2\over {r^2+r^{-2}}}
[-(r-r^{-1}) \de^{b_1}_{b_2} \de^{a_1}_{c_1} \de^{c_2}_{a_2}
+\CC{a_1}{a_2}{b_1}{b_2}{c_1}{c_2}]\\
& &\cc{a_1}{a_2}{0}{b_2}{0}{c_2}={2\over {r^2+r^{-2}}}
\CC{a_1}{a_2}{0}{b_2}{0}{c_2}\\
& &\cc{0}{a_2}{b_1}{b_2}{0}{c_2}={2\over {r^2+r^{-2}}}
[-(r-r^{-1})\de^{b_1}_{b_2} \de^{c_2}_{a_2}+
\CC{0}{a_2}{b_1}{b_2}{0}{c_2}]
\ena
The structure constants $\Cb$ (appearing in the $q$-Lie
algebra of $\IGLqrN$, see later) are given by
\eqa
& &\CC{c_1}{c_2}{b_1}{b_2}{d_1}{d_2}=
\rinv [-\de^{b_1}_{b_2} \de^{c_1}_{d_1}
 \de^{d_2}_{c_2}+\LL{a}{a}{d_1}{d_2}{c_1}{c_2}{b_1}{b_2}]
 \label{C1}\\
& & \mbox{~~~~~~~~~~~~~~= structure constants of $\GLqrN$}
 \nonumber \\
& &\CC{c_1}{c_2}{0}{b_2}{0}{d_2}=-{q_{0c_2}\over q_{0c_1}}
\R{c_1d_2}{b_2c_2} \label{C2}\\
& &\CC{0}{c_2}{b_1}{b_2}{0}{d_2}=
\rinv [-\de^{b_1}_{b_2}
 \de^{d_2}_{c_2}+ d^{f_2} d^{-1}_{c_2} \R{f_2b_1}{c_2a}
 \R{ad_2}{b_2f_2}]
\label{C3}
\ena

We conclude this Section by checking that
the functionals $f$ and $\chi$
in (\ref{fin}) and (\ref{chiin})
close on the algebra
(\ref{bico1}), (\ref{bico2})-(\ref{bico4}), where
the product of functionals is defined by the coproduct
$\D$ in $\IGLqrN$. This result is expected,
since the functionals in (\ref{fin}) and (\ref{chiin})
correspond to a
bicovariant differential
calculus on $\IGLqrN$.

To prove this, we first note that in
$\GLqrNo$ the subset in (\ref{fin}) and (\ref{chiin})
closes by itself on the bicovariant
algebra (\ref{bico1}), (\ref{bico2})-(\ref{bico4}).
This is due to the particular
index structure of the tensors $\Cb$ and $\Lambda$, and to
the vanishing of the $f$ components in (\ref{fzero}).
The nonvanishing components of
$\Cb$ and $\Lambda$ that enter the operatorial
bicovariance  conditions (where the free adjoint indices
are taken as $ab, 0b$), are
given in (\ref{C1})-(\ref{C3}) and (\ref{L1})-(\ref{L4}).
Finally, we know that the $f$ functionals
vanish on $KerP$, and so do the $\chi$ functionals
(as can be deduced from their definition in terms
of the $f$ functionals, eq. (\ref{defchi})).
{} From the discussion after Theorem 1
it follows that they are well defined on $\IGLqrN$, and
that their products involve the $\IGLqrN$ coproduct
$\D$.
\sk
Thus the  relations (\ref{bico1}), (\ref{bico2})-(\ref{bico4})
hold for the functionals (\ref{fin}) and (\ref{chiin})
on $\IGLqrN$. They are the bicovariance conditions
corresponding to a consistent differential calculus on $\IGLqrN$.

\sk
Using the values of the $\Lambda$ and $\Cb$
tensors in (\ref{L1})-(\ref{L4}) and (\ref{C1})-(\ref{C3}),
we can explicitly write the ``$q$-Lie algebra"
of $\IGLqrN$  as:
\eq
 \cchi{c_1}{c_2}\cchi{b_1}{b_2}-
\LL{a_1}{a_2}{d_1}{d_2}{c_1}{c_2}{b_1}{b_2}~
\cchi{a_1}{a_2} \cchi{d_1}{d_2}=\rinv [-\de^{b_1}_{b_2}
\de^{c_1}_{d_1} \de^{d_2}_{c_2} +
\LL{a}{a}{d_1}{d_2}{c_1}{c_2}{b_1}
{b_2}] \cchi{d_1}{d_2}\label{qLie1}
\en
\eqa
& &\cchi{c_1}{c_2} \cchi{0}{b_2} + (r-r^{-1}) {q_{0c_2}\over
q_{0a_2}} \R{a_2d_2}{b_2c_2} \cchi{c_1}{a_2} \cchi{0}{d_2}
- ~~~~~~~~~~~~~~~~~~~~~~~~~~~~~~~~~~~~~~~~~~~~~~~
{}~~~~~\nonumber\\
& &~~~~~~~~~~~~~~~~~~~~~
 -{q_{0c_2}\over q_{0c_1}} \Rinv{a_2c_1}{g_2d_1}
\R{g_2d_2}{b_2c_2} \cchi{0}{a_2} \cchi{d_1}{d_2}
=-{q_{0c_2}\over q_{0c_1}} \R{c_1d_2}{b_2c_2} ~\cchi{0}{d_2}
\label{qLie2}\\
& &\cchi{0}{c_2} \cchi{b_1}{b_2}-{q_{0a_1}\over q_{0a_2}}
d^{f_2} d^{-1}_{c_2} \R{f_2b_1}{c_2a_1} \R{a_2d_2}{b_2f_2}
\cchi{a_1}{a_2} \cchi{0}{d_2}=\nonumber\\
& &~~~~~~~~~~~~~~~~~~~~~
\rinv [-\de^{b_1}_{b_2} \de^{d_2}_{c_2} +d^{f_2} d^{-1}_{c_2}
\R{f_2b_1}{c_2a}
\R{ad_2}{b_2f_2} ] \cchi{0}{d_2} \label{qLie3}\\
& &\cchi{0}{c_2}\cchi{0}{b_2}- {q_{0c_2}\over
q_{0a_2}} r^{-1}~\R{a_2d_2}{b_2c_2}
{}~\cchi{0}{a_2} \cchi{0}{d_2}=0 \label{qLie4}
\ena
\noi where $\LL{a_1}{a_2}{d_1}{d_2}{c_1}{c_2}{b_1}{b_2}$
is the
braiding matrix
of $GL_q(n)$, given in (\ref{L1}), so that the commutations in
(\ref{qLie1}) are those of the $q$-subalgebra $GL_q(n)$.
Note that the $\rone$ limit on the right hand sides
of (\ref{qLie1})
and (\ref{qLie3}) is finite, since the terms
in square parentheses
are a (finite) series in $r-r^{-1}$ whose $0-th$
order part vanishes
(see \cite{Aschieri1}, eq. (5.55)).

\sk
%%%%%%%%%%%%%%%%%%%%%%%%%%%

\sect{The multiparametric quantum plane as a $q$-coset space}

%%%%%%%%%%%%%%%%%%%%%%%%%%%

In this Section we derive the differential calculus
on the quantum plane
\eq
{\IGLqrN \over \GLqrN},
\en
\noi i.e. the subalgebra
generated by the coordinates $x^a$. The coordinates $x^a$
satisfy the commutations
(\ref{RTTIGL3}):
\eq
\R{ab}{ef} x^e x^f=r x^b x^a
\en
\noi The main difference with the more conventional approach
to the quantum plane is that now the coordinates do not trivially
commute with the $\GLqrN$ $q$-group elements, but
$q$-commute according to relations (\ref{RTTIGL2}):
\eq
\R{ab}{ef} \T{e}{c} x^f= {q_{0c} \over r} x^b \T{a}{c}
\label{RTTIGL2bis}
\en
\noi {\sl Lemma:} $\cchi{b}{c} (a)=0$ when a is a
polynomial in $x^a$ and $u$ with all monomials
containing at least one $x^a$. This is easily proved
by observing that no tensor exists with the correct
index structure.  For $s=1$ we can extend this lemma
even to  $u \cdots u$, since
for example
\eq
\cchi{b}{c} (u)={{s-1}\over {r-\rminus}} \de^b_c
\en
\noi and using the coproduct rule (\ref{copchi1})
 one finds that
$\cchi{b}{c}(u \cdots u)$
is always proportional to
$s-1$.
\sk
\noi {\sl Theorem:} $\cchi{b}{c} * a=0$ when $a$ is
a polynomial in $x^a$ and $s=1$.
\sk
\noi {\sl Proof:} we have $\cchi{b}{c} * a=(id\otimes
\cchi{b}{c})(a_1\otimes
a_2)=a_1 \cchi{b}{c} (a_2)$. We use here the standard notation
$\D (a) \equiv a_1 \otimes a_2$.  Since $a_2$ is a polynomial
in $x^a$ and $u$ (use the coproduct rule (\ref{Dx})), and
$\cchi{b}{c}$ vanishes on such a polynomial when $s=1$
(previous Lemma),
the theorem is proved. Q.E.D.
\sk
Because of this theorem we will henceforth set $s=1$: then
we can write the exterior derivative of an element
of the quantum plane as
\eq
da= (\chi_s*a) V^s \label{daflat}
\en
\noi (with $\chi_s \equiv \cchi{0}{s}$, $V^s \equiv \ome{0}{s}$),
i.e. only in terms of the ``q-vielbein" $V^s$.
\sk
The action and value of $\chi_s$ on the coordinates
is easily computed, cf. the
definition in  (\ref{chiin}):
\eq
\chi_s * x^a=-{r\over q_{0s}} \T{a}{s},~~~\chi_s (x^a)=-{r\over
q_{0s}} \de^a_s
\en
\noi so that the exterior derivative of $x^a$ is:
\eq
dx^a=-{r\over q_{0s}} \T{a}{s} V^s \label{dxa}
\en
\noi and gives the relation between the $q$-vielbein $V^s$
and the differentials $dx^a$.
\sk
\noi {\sl Theorem:}
The Leibniz rule for the ``$q$-partial derivatives" $\chi_c$
is given by :
\eq
\chi_c * (ab)=(\chi_d * a) \f{d}{c} * b + a \chi_c * b
\label{Leibnizplane}
\en
\noi where $\f{d}{c} \equiv \ff{0}{d0}{c}$.
\sk
\noi {\sl Proof:} compute the left-hand side using the
coproduct (\ref{copchi2}):
\eqa
& &\chi_c * (ab)=(id \otimes \chi_c)
(a_1 b_1 \otimes a_2 b_2)=a_1b_1 (\chi_d \otimes \f{d}{c} +
\epsi\otimes \chi_c)(a_2 \otimes b_2)=\nonumber\\
& & \chi_d(a_2)a_1b_1 \f{d}{c}(b_2)+a_1 \epsi(a_2)b_1 \chi_c(b_2)
\ena
\noi This is easily seen to coincide with the right-hand side
of (\ref{Leibnizplane}), if one
remembers that $a_1 \epsi(a_2)$=$a$ in virtue of (\ref{axiom2}).
Q.E.D.
\sk
The $x^a$ and $V^b$ $q$-commute as (cf. (\ref{Vxcomm})):
\eq
V^a x^b= (q_{0a})^{-1} x^b V^a
\en
\noi and via eq. (\ref{dxa}) and (\ref{RTTIGL2bis})
 we find the $dx^a, x^b$commutations :
\eq
dx^a x^b=r^{-1} \Rinv{ab}{ef} x^f dx^e
\en
\noi After acting on this equation with $d$ we obtain:
\eq
dx^a \we dx^b= -r^{-1} \Rinv{ab}{ef} dx^f \we dx^e
\en
\noi which reproduce the known commutations between
the differentials
of the quantum plane, cf. ref. \cite{qplane}.
\sk
The commutations between the partial derivatives are given in
eq.(\ref{qLie4}).
\sk
As in \cite{qplane}, all the relations of this Section
are covariant under the
$\GLqrN$ action:
\eq
x^a \longrightarrow \T{a}{b} \otimes x^b
\en
\noi {\sl Note: }the partial derivatives $\chi_c$,
and in general all the tangent vectors $\chi$
of this paper have "flat" indices. To compare
them with the partial derivatives discussed in
\cite{qplane}, which have "curved" indices,
we need to define the functionals $\chit_s$:
\eq
\chit_s * a \equiv - {q_{0a}\over r} (\chi_a * a)
\kappa (\T{a}{s})
\en
\noi whose action and value on the coordinates is
\eq
\chit_s * x^a=\de^a_s I,~~~\chit_s (x^a)=
\epsi(\chit_s * x^a)=\de^a_s
\en
\noi so that
\eq
da=(\chit_s * a) ~dx^s
\en
\noi which is equation (\ref{daflat}) in ``curved" indices
(Note: ref. \cite{qplane} adopts a definition of $\chit_s$ such that
$da= dx^s~(\chit_s * a)$).
\sk
The results of this Section are applied to the
multiparametric quantum plane  $IGL_{qr}(2)/GL_{qr}(2)$
at the end of the Table. The usual relations
of the uniparametric case \cite{qplane} are recovered
after setting $q=r$.
\vskip 1cm
\leftline{\bf Acknowledgements}
\sk
It is a pleasure to acknowledge valuable discussions with
C. De Concini,  C. Chryssomalakos, B. Jur\v co, P. Kulish, P. Schupp
and M. Tarlini.
\vfill\eject

%%%%%%%%%%%%%%%%%%%%%%%%%%%%%%%%%%%%%%%%%

\app{ The Hopf algebra axioms}

%%%%%%%%%%%%%%%%%%%%%%%%%%%%%%%%%%%%%%%%%

A Hopf algebra over the field $K$ is a
unital algebra over $K$ endowed
with the  linear maps:
\eqa
& &\D~: ~~A \rightarrow A\otimes A\\
& &\epsi~:~~ A\rightarrow K\\
& &\kappa~:~~A\rightarrow A
\ena
\noi satisfying the following
properties $\forall a,b \in A$:
\eq
(\D \otimes id)\D(a) = (id \otimes \D)\D(a)
\en
\eq
(\epsi \otimes id)\D(a)=(id \otimes \epsi)\D(a) =a \label{axiom2}
\en
\eq
m(\kappa\otimes id)\D(a)=m(id \otimes \kappa)\D(a)
=\epsi(a)I \label{kappadelta}
\en
\eq
\D(ab)=\D(a)\D(b)~;~~\D(I)=I\otimes I
\en
\eq
\epsi(ab)=\epsi(a)\epsi(b)~;~~\epsi(I)=1
\en
\noi where $m$ is the multiplication map $m(a\otimes b)
= ab$.
{} From  these axioms we deduce:
\eq
\kappa(ab)=\kappa(b)\kappa(a)~;~~\D[\kappa(a)]=\tau(\kappa\otimes
\kappa)\D(a)~;~~\epsi[\kappa(a)]=\epsi(a)~;~~\kappa(I)=I
\en
where $\tau(a \otimes b)=b\otimes a$ is the twist map.
\vfill\eject

%%%%%%%%%%%%%%%%%%%%%%%%%%%%%%%%%%%%%

\centerline{{\bf Table}}

%%%%%%%%%%%%%%%%%%%%%%%%%%%%%%%%%%%%%

\sk
\centerline{The quantum group $IGL_{q,r}(2)$ and its differential
calculus}
\sk
\sk
\noi {\sl Parameters:} $q (\equiv q_{12}), q_{01}, q_{02}, r$
\sk
\noi {\sl $R$ and $D$-matrices of $GL_{q,r}(2)$:}
\[
\R{ab}{cd}=\left(
\begin{array}{cccc}
r & 0 & 0 & 0 \\
                                 0 & {r\over q} & 0 & 0 \\
                                 0 & r-\rminus & {q\over r} & 0 \\
                                 0 & 0 & 0 & r \end{array} \right)
,~~D^a_{~b}=\left( \begin{array}{cc} r & 0  \\ 0 & r^3 \end{array}
\right)
\]
\sk
\noi {\sl $\T{A}{B}$ (A,B=0,1,2):
fundamental representation of $IGL_{q,r}(2)$}
\sk
\[
\T{A}{B}=\left( \begin{array}{ccc}  u & x^1 & x^2 \\
                                      0  & \al & \be \\
                                      0 & \ga & \de
 \end{array} \right)
\]
\sk
\noi {\sl Determinant of  $IGL_{q,r}(2)$ and definition of $\xi$}
\sk
\[
\det \T{A}{B}=u \det \T{a}{b},  ~~\mbox{where}~
\det \T{a}{b}=\al\de-{r^2\over q} \be\ga
\]
\[
\xi \det \T{A}{B}=\det \T{A}{B} \xi=I
\]
\sk
\noi {\sl Basis elements generating $IGL_{q,r}(2)$}
\sk
\[
\al, \be , \ga , \de , ~x^1,~x^2, ~u, ~\xi
\]
\sk
\noi {\sl Commutations of the basis elements }
\sk
\[
\al\be={r^2\over q}\be\al,~~\al\ga=q\ga\al,~~\be\de=q\de\be,
{}~~\ga\de={r^2\over q}\de\ga
\]
\[
\be\ga={q^2\over r^2}\ga\be,~~\al\de-\de\al=
{r\over q}(r-\rminus)\be\ga,
\]
\[
\al x^1={q_{01}\over r^2} x^1 \al,
 ~~~\be x^1={q_{02}\over r^2} x^1 \be,
\]
\[
\al x^2=q{q_{01}\over r^2} x^2 \al,
{}~~~\be x^2=q{q_{02}\over r^2} x^2 \be,
\]
\[
\ga x^1={q_{01}\over q} x^1
\ga-{r\over q} (r-\rminus) \al x^2, ~~~
\de x^1={q_{02}\over q} x^1
 \de-{r\over q} (r-\rminus) \be x^2,
\]
\[
\ga x^2={q_{01}\over r^2} x^2 \ga,~~~
\de x^2={q_{02}\over r^2} x^2 \de
\]
\[
x^1 x^2=q x^2 x^1
\]
\[
\T{a}{b} u={q_{0b}\over q_{0a}} u \T{a}{b},
{}~~~x^a u= (q_{0a})^{-1} u x^a
\]
\[
(\det \T{A}{B})\T{A}{B} ={q_{0A} q_{1A} q_{2A}
 \over q_{0B} q_{1B} q_{2B}}
 \T{A}{B} (\det \T{A}{B}), ~q_{AA} \equiv r, ~q_{AB}
\equiv {r^2\over q_{BA}}
\]
\[
\T{A}{B} \xi={q_{0A} q_{1A} q_{2A} \over q_{0B} q_{1B} q_{2B}}
\xi \T{A}{B}
\]
\sk
\noi {\sl Conditions for centrality of
$\det \T{A}{B}=u \det \T{a}{b}$, $\det \T{a}{b}$ and $u$}
\sk
centrality of $u \det \T{a}{b}$ ~$\Longleftrightarrow$
$~~q_{01} q_{02}=r^2, ~q_{01}=q$

centrality of $\det \T{a}{b}$ ~~~$\Longleftrightarrow$
$~~q_{01} q_{02}=r^2, ~q=r$

centrality of $u$ ~~~~~~~~~~~$\Longleftrightarrow$
$~~q_{01}= q_{02}=1$
\sk\sk

\noi {\sl Inverse of $\T{A}{B}$}

\[
\Ti{A}{B}=\left(
\begin{array}{cc}  \det \T{a}{b} \xi & -\Ti{a}{b} x^b
                                    \det \T{a}{b} \xi  \\
                                      0  & \Ti{a}{b}
                 \end{array} \right)
\]
\[
\Ti{a}{b}= \xi u \left( \begin{array}{cc} \de & -\qm \be  \\
                                -q\ga      & \al
                 \end{array} \right)
\]
\sk

\noi {\sl Commutations of the left-invariant one-forms $\om$}
\sk
Notations: $\om^1 \equiv \ome{1}{1},\om^+
\equiv \ome{1}{2},\om^-\equiv \ome{2}{1},
\om^2 \equiv \ome{2}{2},~V^1\equiv \ome{0}{1},
V^2\equiv \ome{0}{2}$

\[\om^1 \we \om^+ + \om^+ \we \om^1 = 0\]
\[\om^1 \we \om^- + \om^- \we \om^1 = 0\]
\[\om^1 \we \om^2 + \om^2 \we \om^1 = (1-r^2)\om^+ \we \om^-
\]
\[\om^+ \we \om^- + \om^- \we \om^+ = 0\]
\[\om^2 \we \om^+ + r^2 \om^+ \we \om^2 =
 r^2 (r^2 - 1)\om^+ \we \om^1
\]
\[\om^2 \we \om^- + r^{-2} \om^- \we \om^2 =
 (1-r^2)\om^- \we \om^1\]
\[\om^2 \we \om^2 =(r^2 - 1)\om^+ \we \om^-
\]
\[\om^1 \we \om^1 = \om^+ \we \om^+ = \om^- \we \om^-=0\]
\sk
\[ \om^1 \we V^1+r^2 V^1 \we \om^1=0\]
\[ \qm {q_{02} \over q_{01}} \om^+ \we V^1 + V^1 \we \om^+=
(1-r^{-2}) \om^1 \we V^2 \]
\[ \om^- \we V^1 + {r^2 \over q}{q_{02}
\over q_{01}} V^1 \we \om^-=0 \]
\[ \om^2 \we V^1 + V^1 \we \om^2 =(1-r^{-2}) q {q_{01}\over q_{02}}
\om^- V^2 \]
\[ \om^1 \we V^2+ V^2 \we \om^1=0\]
\[ \qm {q_{02} \over q_{01}}\om^+ \we V^2+V^2 \we \om^+=0\]
\[ {q\over r^2}{q_{01} \over q_{02}}\om^- \we V^2+V^2 \we \om^-=
(1-r^2) V^1 \we  \om^1 \]
\[ \om^2 \we  V^2+r^2 \we V^2 =(r^2-1)[(1-r^2)\om^1 \we V^2+
{r^2 \over q}{q_{02} \over q_{01}} \om^+ \we V^1] \]

\noi {\sl Cartan-Maurer equations:}
\sk
\[ d\om^1+r\om^+ \we \om^-=0 \]
\[ d\om^+ + r \om^+(-r^2 \om^1 + \om^2)=0 \]
\[ d\om^- + r (-r^2 \om^1 + \om^2)\om^-=0 \]
\[ d\om^2 - r \om^+ \we \om^-=0 \]
\[ dV^1-{q\over r}{q_{01} \over q_{02}}\om^-\we V^2-\rminus
\om^1 \we V^1 =0\]
\[ dV^2-{r\over q}{q_{02} \over q_{01}}\om^+ \we V^1-
\rminus \om^2 \we V^2-(r-\rminus) V^2 \we \om^1=0\]
\sk

\noi {\sl The q-Lie algebra:}
\sk
Notations: $\chi^1 \equiv \cchi{1}{1},\chi^+
\equiv \cchi{1}{2},\chi^-\equiv \cchi{2}{1},
\chi^2 \equiv \cchi{2},~P^1\equiv \cchi{0}{1},
P^2\equiv \cchi{0}{2}$

\[ \chi_1 \chi_+ - \chi_+ \chi_1 -(r^4-r^2)\chi_2 \chi_+ =
 r^3 \chi_+\]
\[ \chi_1 \chi_- - \chi_- \chi_1 +
(r^2-1)\chi_2 \chi_- = -r \chi_-\]
\[ \chi_1 \chi_2 - \chi_2 \chi_1 =0\]
\[ \chi_+ \chi_- - \chi_- \chi_+ + (1-r^2) \chi_2 \chi_1-(1-r^2)
\chi_2 \chi_2 = r (\chi_1-\chi_2)
\]
\[ \chi_+ \chi_2-r^2 \chi_2 \chi_+ = r\chi_+\]
\[ \chi_- \chi_2-r^{-2} \chi_2 \chi_- = -\rminus \chi_-\]
\[ r^2 \chi_1 \P_1-\P_1 \chi_1+(r^2-1) \P_2 \chi_-=-r\P_1 \]
\[ q{q_{01} \over q_{02}} \chi_+ \P_1-\P_1 \chi_+-r^2 (1-r^2)
   \chi_2 \P_2=r^3 \P_2 \]
\[ \chi_- \P_1-{q\over r^2}{q_{01} \over q_{02}}\P_1 \chi_-=0 \]
\[ \chi_2 \P_1-\P_1 \chi_2=0 \]
\[ \chi_1 \P_2-\P_2 \chi_1 + (r^2-1)
{q\over r^2}{q_{01} \over q_{02}}
    \chi_+ \P_1=0 \]
\[ \chi_+ \P_2 - \qm {q_{02} \over q_{01}}\P_2 \chi_+=0 \]
\[  {r^2\over q}{q_{02} \over q_{01}}\chi_- \P_2- \P_2 \chi_-
+(1-r^2) \chi_2 \P_1=-r \P_1 \]
\[ r^2 \chi_2 \P_2-\P_2 \chi_2=-r\P_2 \]
\sk
\[ \P_1 \P_2-{q\over r^2}{q_{01} \over q_{02}}\P_2 \P_1=0 \]
\sk

\noi {\sl The exterior derivative of the basis elements}
\sk
\[
\begin{array}{l}
 d\alpha={{s-r^2}\over{r^3-q}}\al\om^1-s
{r \over q_{12}}\beta\om^++{{s-1}\over{r-
\rminus}}\al\om^2\\
 d\beta={{-r^2+s(1-r^2+r^4)}\over{r^3-r}}
\beta\om^1-s {q_{12}\over r} \alpha\om^-
+{{s-r^2}\over{r^3-r}}\be\om^2\\
d\gamma={{s-r^2}\over{r^3-r}}\ga\om^1-s
{r\over q_{12}}\de\om^+
+{{s-1}\over{r-\rminus}}\ga\om^2\\
d\de={{-r^2+s(1-r^2+r^4)}\over{r^3-r}}
\de\om^1-s {q_{12}\over r} \ga\om^-+
{{s-r^2}\over{r^3-r}}\de\om^2\\
dx^1=-{s r\over q_{01}} \al V^1-{s r\over q_{02}} \be V^2+
{{s-1}\over {r-\rminus}} x^1 \tau\\
dx^2=-{s r\over q_{01}} \ga V^1-{s r\over q_{02}} \de V^2+
{{s-1}\over {r-\rminus}} x^2 \tau\\
du={{s-1}\over {r-\rminus}} u \tau\\
d\xi={{r^2 s^{-N-1} -1}\over {r-\rminus}} \xi \tau,~~d\zeta=
{{r^2 s^{-N} -1}\over {r-\rminus}} \zeta \tau\\
d(\det \T{A}{B})={{r^{-2} s^{N+1} -1}
\over {r-\rminus}} (\det \T{A}{B}) \tau,
{}~~d(\det \T{a}{b})=
{{r^{-2 }s^{N} -1}\over {r-\rminus}}  (\det \T{A}{B})\tau
\end{array}
\]
\noi {\sl The $\om^i$ in
terms of the exterior derivatives on $\al,\be,\ga,
\de, x^1, x^2 ,u$:}
\[
\begin{array}{l}
 \om^1={r\over{s(-r^2-r^4+s+sr^4)}}
[(r^2-s)(\kappa (\al)da+\kappa
(\be)d\ga)+r^2(s-1)(\kappa (\ga)d\be + \kappa (\de) d\de)]\\
 \om^+=-{1\over s} {q_{12}\over r}[\kappa (\ga) d\al +
\kappa (\de) d\ga]\\
\om^-=-{1\over s}{r\over {q_{12}}} [\kappa (\al) d\be +
\kappa (\be) d\de]\\
\om^2={r\over{s(-r^2-r^4+s+sr^4)}}
[(s-r^2-sr^2+sr^4)(\kappa (\al)
d\al+\kappa (\be) d\ga)+(r^2-s)(\kappa
(\ga) d\be+\kappa (\de) d\de)]\\
V^1=-{q_{01}\over{s r}} [\kappa (\al)dx^1+\kappa
 (\be)dx^2+\kappa (x^1)du]\\
V^2=-{q_{02}\over {s r}} [\kappa (\ga)dx^1+\kappa
 (\de)dx^2+\kappa (x^2)du]
\end{array}
\]
\sk
\noi {\sl The multiparametric quantum plane
$IGL_{qr}(2)/GL_{qr}(2)$}
\[
\begin{array}{l}
x^1 x^2=q x^2 x^1\\
{}\\
dx^1 x^1=r^{-2} x^1 dx^1\\
dx^1~ x^2= {q\over r^2} x^2 dx^1\\
dx^2~ x^1=(r^{-2}-1) x^2 dx^1+\qm x^1 dx^2\\
dx^2 x^2=r^{-2} x^2 dx^2\\
{}\\
dx^1 \we dx^2= -{q\over r^2}dx^2 \we dx^1
\end{array}
\]
\vfill\eject

\vfill\eject
\end{document}